\documentclass[twocolumn, dvipsnames, table]{aastex631}
\usepackage{graphicx} 
\usepackage{amsmath,bm}
\usepackage{mathtools}   
\usepackage{graphicx}
\usepackage{verbatim}
\usepackage{comment}
\usepackage{changepage}
\usepackage{float}
\usepackage{hyperref}
\usepackage{booktabs}

\newcommand{\bfm}[1]{\mathbf{#1}}
\newcommand{\dd}{\rm d}

\newcommand{\curl}[1]{\mathbf{\nabla} \times \mathbf{#1}}
\newcommand{\dvg}[1]{\mathbf{\nabla} \cdot \mathbf{#1}}
\newcommand{\Hquad}{\hspace{0.5em}}

\renewcommand{\arraystretch}{1.1}

\begin{document}
\title{Nonlinear Decay of Fast Magnetosonic Waves through Weak Turbulence: Force-Free Electrodynamics Simulations}
\author[0000-0001-6541-734X]{Siddhant Solanki}
\affiliation{Department of Astronomy, University of Maryland, 7901 Regents Drive, College Park, MD 20742, USA} 
\author[0000-0002-5349-7116]{Jens F. Mahlmann}
\affiliation{Department of Physics \& Astronomy, Wilder Laboratory, Dartmouth College, Hanover, NH 03755, USA}
\author[0000-0001-7801-0362]{Alexander Philippov}
\affiliation{Department of Physics, University of Maryland, 7901 Regents Drive, College Park, MD 20742, USA}
\affiliation{Physics Department, Stanford University, 382 Via Pueblo Mall, Stanford, CA 94305, USA}
\affiliation{
Kavli Institute for Particle Astrophysics and Cosmology, Physics and Astrophysics Building, 452 Lomita Mall, Stanford University, Stanford, CA 94305-
4085, USA}

\begin{abstract}
We investigate the propagation of low-frequency fast-magnetosonic (FMS) waves in highly magnetized environments. Such conditions are relevant to the escape of GHz fast radio bursts potentially produced in the inner magnetospheres of magnetars. It remains an open question whether such waves can escape without substantial reprocessing. Using relativistic force-free electrodynamics simulations, we confirm the key theoretical predictions of \citet{golbraikh_2023ApJ...957..102G} and demonstrate that FMS waves undergo efficient nonlinear conversion into secondary FMS and Alfvén waves via the parametric decay instability. This process continues to drain energy from the primary FMS waves even after approximate energy equipartition between the FMS and Alfvén components is established. The resulting spectrum of excited waves is broad, extending across much of the inertial range in $k$-space within the simulation domain. Our results indicate that FMS waves likely do not escape magnetar magnetospheres without substantial dissipation and spectral broadening.

\end{abstract}

\section{Introduction} \label{section:Introduction}
In the magnetospheres of isolated neutron stars, the electromagnetic (EM) energy density exceeds the plasma's thermal and rest-mass energy density by many orders of magnitude \citep{goldrecih_julian_1969ApJ...157..869G}. Such environments can be modeled using force-free electrodynamics \citep[FFE,][]{gruzinov_force_free_1999astro.ph..2288G,blandford_to_the_lighthouse_2002luml.conf..381B}. In this regime, modes with frequencies $\omega < \omega_p$, where $\omega_p$ is the local plasma frequency, behave like magnetohydrodynamic (MHD) waves. FFE supports two linear modes: fast magnetosonic (FMS) and Alfvén waves. When the wave amplitudes are small compared to the background field, AWs propagate along magnetic field lines and remain confined, whereas FMS waves can travel across the field lines and potentially escape the system.

The different behavior of FMS and Alfvén waves constrains theoretical models of Fast Radio Bursts (FRBs). While the emission mechanism of FRBs is still unknown, some bursts are likely powered by magnetars. So-called `near-field' models associate the FRB signal generation with the inner magnetar magnetosphere, at distances of $10$--$10^2 R_{\rm NS}$ from the star, where $R_{\rm NS} \sim 10^6$ cm is the magnetar radius \citep{Kumar2020:2004.00644v1,Lu_kumar_frb_2020MNRAS.498.1397L,Zhang_IC2021:2111.06571v1,zhang_review_2023RvMP...95c5005Z,Qu2024:2404.11948v2}. At these distances, for plasma densities expected in magnetar magnetospheres \citep{beloborodov_multiplicity_2013ApJ...762...13B}, the plasma frequency greatly exceeds $1\,\mathrm{GHz}$, and the wave amplitude remains below the background magnetic field. Thus, propagating FRBs correspond to FMS waves of FFE \citep{golbraikh_2023ApJ...957..102G}.

As these waves travel in an FFE background, the local electric currents adjust to preserve the orthogonality of electric, $\mathbf{E}$, and magnetic, $\mathbf{B}$, fields, $\mathbf{E}\cdot\mathbf{B}=0$, introducing nonlinearity to the system \citep{McKinney2006,Komissarov2011}. The nonlinearity allows for resonant interactions: linear waves can transform into lower-frequency modes (parametric decay) or interact with other linear waves to generate higher-frequency modes \citep{Zakharov1992}. This process is commonly described as weak turbulence (WT). 

Understanding WT in the relativistic force-free regime is critical for evaluating FRB escape from the inner magnetar magnetosphere. As emphasized by \cite{lyubarsky_review_frb_2021Univ....7...56L}, WT may quench FRB escape: the FMS waves can resonantly transfer their energy to AWs that become trapped along closed field lines. To quantitatively evaluate this proposal, \cite{golbraikh_2023ApJ...957..102G} solved WT spectral equations constructed in the random phase approximation \citep[see the non-relativistic framework by][]{chandran_2005PhRvL..95z5004C}. Their treatment includes the bidirectional processes $F \leftrightarrow F + A$ and $F \leftrightarrow A + A$, where $F$ and $A$ correspond to FMS and Alfvén waves, respectively. These processes, and their respective inverses, represent the conversion of a high-frequency FMS wave into a lower-frequency FMS wave and an AW, or two AWs. However, the calculations by \cite{golbraikh_2023ApJ...957..102G} do not yet account for potential interactions between two counter-propagating AWs that could lead to weak Alfvénic turbulence \citep{2012PhRvE..85c6406S,Howes_2013PhPl...20g2302H,alfven_turbulence_tenbarge_2021JPlPh..87f9014T,Bart_2021JPlPh..87e9012R}. Once a significant fraction of energy is deposited into AWs from the $F \leftrightarrow F + A$ and $F \leftrightarrow A + A$ processes, weak Alfvénic turbulence can become important. In this process, counter-propagating AWs interact with a 2D condensate composed of nonlinear modes \citep{2012PhRvE..85c6406S}, making the WT treatment of three-wave interactions challenging. Fully nonlinear force-free simulations complement spectral methods and offer additional insight. Moreover, such simulations conserve Poynting flux, which, as we demonstrate explicitly, is carried in nonlinear fields and cannot be fully captured by a WT approach.

In this work, we simulate WT in force-free magnetospheres using three-dimensional high-order, high-resolution FFE simulations. Our method integrates all nonlinear interactions without pre-selecting dominant resonant channels.

\section{Wave Interactions in Magnetar Magnetospheres} \label{section:Theory: Weak Turbulence in Force Free}

As mentioned above, within the inner magnetospheres of magnetars, FRBs correspond to FMS waves of FFE. In the following sections, we discuss how these waves interact and evolve via nonlinear processes.

\subsection{Resonant three-wave interactions} \label{subsection: Interactions of Three Resonant Modes}
Three-wave interactions constitute the lowest-order nonlinear wave-coupling processes and form the foundational elements of WT theory \citep{Zakharov1992}. A resonant interaction among three waves, characterized by wavevectors $\bfm{k}_1$, $\bfm{k}_2$, and $\bfm{k}_3$ and angular frequencies $\omega_1$, $\omega_2$, and $\omega_3$, is permitted when the following wave-matching (resonance) conditions are satisfied:
\begin{align}
    \bfm{k_1} &= \bfm{k_2} + \bfm{k_3} \label{eq:resonance_k}\,, \\
    \omega_1(\bfm{k_1}) &= \omega_2(\bfm{k_2}) + \omega_3(\bfm{k_3}) \label{eq:resonance_omega},
\end{align}
where $\omega_1$, $\omega_2$, and $\omega_3$ are determined by the wave dispersion relations of the medium. If the three waves satisfy Equations~(\ref{eq:resonance_k}) and~(\ref{eq:resonance_omega}), they form a resonant triad. Such resonant interactions require nonvanishing interaction coefficients, which determine the strength of the coupling and depend on the nonlinear properties of the medium supporting the waves.

The FFE dispersion relations for FMS and Alfvén waves in the ultra-magnetized limit are $\omega^F = |\bfm{k}|c$ and $\omega^A = |\bfm{k}\cdot\hat{\bfm{B}}|c = |k_{\parallel}|c$, respectively, where $\hat{\bfm{B}}$ is the unit vector along the background magnetic field. In the most general case, FFE permits the interactions $F \leftrightarrow F + A$, $F \leftrightarrow A + A$, and $A \leftrightarrow A + A$, while the $F \leftrightarrow F + F$ and $A \leftrightarrow A + F$ interactions are forbidden \citep{thompson_blaes_1998PhRvD..57.3219T,Li_beloborodov_2019ApJ...881...13L}. In Appendix~\ref{appendix:three_wave_interactions}, we compute the interaction rates $V$ for the $F \leftrightarrow F + A$ and $F \leftrightarrow A + A$ processes, denoted by $V_{FFA}$ and $V_{FAA}$, respectively, which scale as $V_{FFA} \sim V_{FAA} \sim V \sim \mathcal{O}(\theta)\,\omega^{3/2} B_0^{-1}$, where $\mathcal{O}(\theta)$ is an order-unity factor depending on the angle between $\bfm{k}_1$ and $\bfm{k}_2$, and $B_0$ is the background magnetic field (see also the appendices of \citealt{Lyubarsky_crab_2019MNRAS.483.1731L,Li_beloborodov_2019ApJ...881...13L}). For an FMS wave with wavevector $\bfm{k}_1$ decaying into product modes with wavevectors $\bfm{k}_2$ and $\bfm{k}_3$, the corresponding matrix coefficients $V_{FFA}$ and $V_{FAA}$ are given by:
\begin{align}
    & V_{FFA}(\bfm{k_1}, \bfm{k_2}, \bfm{k_3}) = \frac{k_{3,\perp}   c^{3/2}}{B_0} \sqrt{\frac{\pi}{2} \frac{|k_{3,\parallel}| k_2}{k_1}}
     \nonumber \\ 
    &\times\left(1 - \mathrm{sign}(k_{3,\parallel}) \frac{k_{2,\parallel}}{k_2} \right) \bfm{\hat{z}} \cdot\left[\bfm{\hat{k}}_{1,\perp} \times \bfm{\hat{k}}_{3,\perp}\right] \label{eq:vffa} \\
    &V_{FAA}(\bfm{k_1}, \bfm{k_2}, \bfm{k_3}) = \frac{k_{3,\perp} c^{3/2}}{B_0} \sqrt{\frac{\pi}{2} \frac{|k_{2,\parallel}| |k_{3,\parallel}|}{k_1}} \nonumber \\ 
    &\times\left(1 - \mathrm{sign}(k_{2,\parallel} k_{3,\parallel}) \right) 
     \bfm{\hat{k}}_{1,\perp} \cdot \left[ \bfm{\hat{k}}_{2,\perp} +  \frac{k_{2,\perp}}{k_{3,\perp}}\bfm{\hat{k}}_{3,\perp}\right] \label{eq:vfaa},
\end{align}
Here, $\bfm{k}_3$ denotes the wavevector of the product Alfvén wave in the $F \leftrightarrow F + A$ process, and the wavevectors $\bfm{k}_i$ satisfy Equation~(\ref{eq:resonance_k}). Subscripts ${\parallel}$ and $\perp$ denote the components of the wavevector parallel and perpendicular to the background magnetic field, respectively.


The $F \leftrightarrow A + A$ process does not constrain the perpendicular components $k_{(2,3),\perp}$ of the excited waves via the resonance conditions (\ref{eq:resonance_k}) and (\ref{eq:resonance_omega}), allowing excitation of modes with $k_{(2,3),\perp} \gg k_{1,\perp}$. This occurs because the frequencies of Alfvén waves are independent of $k_\perp$, and highly anisotropic, counter-propagating Alfvén waves with $k_\perp \gg k_\parallel$ and $\bfm{k_2} \approx -\bfm{k_3}$ satisfy the three-wave resonance as long as the condition $\bfm{k_{2,\perp}} + \bfm{k_{3,\perp}} = \bfm{k_{1,\perp}}$ is fulfilled. Thus, the decay of FMS waves into Alfvén waves with arbitrarily high $k_\perp$ is possible \citep{golbraikh_2023ApJ...957..102G}. Moreover, the rate of the $F \to A + A$ process in this limit does not depend on $\bfm{k_2}$ and $\bfm{k_3}$. Using the wave dispersion relations and the resonance conditions in Equations~(\ref{eq:resonance_k}) and~(\ref{eq:resonance_omega}), and assuming counter-propagating Alfvén waves, one finds
$k_{2,\parallel} = (k_1 + k_{1,\parallel})/{2}$, $k_{3,\parallel} = ({k_1 - k_{1,\parallel}})/{2}$.
Substituting these expressions into Equation~(\ref{eq:vfaa}) and taking $k_{1,\perp}/k_{2,\perp},\, k_{1,\perp}/k_{3,\perp} \ll 1$, the interaction coefficient reduces to:
\begin{align}
    V_{FAA}(k_{2,\perp}, k_{3,\perp} \gg k_{1,\perp}) \approx \sqrt{\frac{\pi}{2}} \frac{(k_1 c)^{3/2} \sin^2\theta }{B_0} \times \nonumber \\ 
    (1 - 2 \cos^2 \Phi),
    \label{eq:vfaa_limit}
\end{align}
Here, $\theta$ is the angle between $\bfm{k}_1$ and $\hat{\bfm{B}}$, and $\Phi$ is the angle between $\bfm{k}_1$ and $\bfm{k}_2$. For a given $\bfm{k}_1$, the maximum value of the interaction coefficient is $\sqrt{{\pi}/{2}} (k_1 c)^{3/2} {\sin^2 \theta}/{B_0},$ which is independent of the ratio $k_{(2,3),\perp} / k_{1,\perp}$.

\begin{table*}[t]
\centering
\caption{Overview of simulation parameters for FMS waves interacting with a tenuous Alfv\'{e}n wave background (top rows, Section~\ref{sec:FWSims}) and mode coupling when the AW and FMS fields have comparable energies (bottom row, Section~\ref{subsec:FMS-Alfven Equipartition}). For each simulation ID, we report the width $\sigma_k$ and the mean wavevector $\bar{\mathbf{k}}$ of the Gaussian wave spectrum. Components of $\bar{\mathbf{k}}$ are given as vectors $(k_x, k_y, k_z)$, with the background magnetic field aligned along the $z$-direction. $U_{\rm FMS} / U_{B_0}$ and $U_{\rm A} / U_{B_0}$ denote the total FMS and Alfv\'{e}n wave energies in units of the background field energy at the start of the simulation.}
\label{tab:sims}
\begin{tabular}{l c c c c l}
\hline\hline
\textbf{Simulation ID} &
$\sigma_k \times [L / 2\pi]$ &
$\bar{\mathbf{k}} \times [L / 2\pi]$ &
$U_{\rm FMS} / U_{B_0}$ &
$U_{\rm A} / U_{B_0}$ &
\textbf{Description} \\
\hline
\texttt{Broad-Strong} & 10   & $(50, 0, 0)$  & $2 \times 10^{-1}$ & $2 \times 10^{-4}$ & \parbox[t]{6cm}{Broad FMS spectrum (comparable wave and background field energies)} \\[6pt]
\texttt{Broad}        & 10   & $(50, 0, 0)$  & $2 \times 10^{-2}$ & $2 \times 10^{-4}$ & \parbox[t]{6cm}{Reference simulation: Broad FMS spectrum} \\[6pt]
\texttt{Broad-Weak}   & 10   & $(50, 0, 0)$  & $2 \times 10^{-3}$ & $2 \times 10^{-4}$ & \parbox[t]{6cm}{Broad FMS spectrum (weakly non-linear regime)} \\[6pt]
\texttt{Narrow}       & 5    & $(50, 0, 0)$  & $2 \times 10^{-2}$ & $2 \times 10^{-4}$ & \parbox[t]{6cm}{Narrow FMS spectrum} \\[6pt]
\texttt{Very-Narrow}  & 1    & $(50, 0, 0)$  & $2 \times 10^{-2}$ & $2 \times 10^{-4}$ & \parbox[t]{6cm}{Very narrow FMS spectrum} \\[6pt]
\texttt{Mono-90}      & 0.01 & $(40, 0, 0)$  & $2 \times 10^{-2}$ & $2 \times 10^{-4}$ & \parbox[t]{6cm}{Monochromatic FMS wave propagating at $90^{\circ}$ with respect to the background field} \\[6pt]
\texttt{Mono-75}      & 0.01 & $(56, 0, 15)$ & $2 \times 10^{-2}$ & $2 \times 10^{-4}$ & \parbox[t]{6cm}{Monochromatic FMS wave propagating at $75^{\circ}$ with respect to the background field} \\[6pt]
\texttt{Mono-60}      & 0.01 & $(52, 0, 30)$ & $2 \times 10^{-2}$ & $2 \times 10^{-4}$ & \parbox[t]{6cm}{Monochromatic FMS wave propagating at $60^{\circ}$ with respect to the background field} \\[6pt]
\texttt{Mono-30}      & 0.01 & $(30, 0, 52)$ & $2 \times 10^{-2}$ & $2 \times 10^{-4}$ & \parbox[t]{6cm}{Monochromatic FMS wave propagating at $30^{\circ}$ with respect to the background field} \\[6pt]
\hline
\texttt{F=A} (FMS)    & 5    & $(30, 30, 30)$ & $2 \times 10^{-2}$ & $2 \times 10^{-2}$ & Energy equipartition \\
\texttt{F=A} (AW1)    & 5    & $(30, 30, 11)$ &                    &                    & \\
\texttt{F=A} (AW2)    & 5    & $(30, 30, 41)$ &                    &                    & \\
\hline\hline
\end{tabular}
\end{table*}

\subsection{Three-wave system} 
\label{subsec:threewaves}

For a set of three resonantly interacting waves (see also Appendix~\ref{appendix:three_wave_interactions}), the evolution equation for the rescaled wave amplitudes, $a_{\bfm{k}_i}$, can be written in the form:
\begin{align}
    \left(\frac{\rm d}{{\rm d}t} + i \omega_1\right) a_{\mathbf k_1} &= i V a_{\mathbf k_2} a_{\mathbf k_3}   \label{eq:three_wave_rescaled_1} \\
    \left(\frac{\rm d}{{\rm d}t}+ i \omega_2\right) a_{\mathbf k_2}  &= i V a_{\mathbf k_1} a_{\mathbf k_3}^* \label{eq:three_wave_rescaled_2} \\
    \left(\frac{\rm d}{{\rm d}t} + i \omega_3\right) a_{\mathbf k_3} &= i V a_{\mathbf k_1} a_{\mathbf k_2}^*,\label{eq:three_wave_rescaled_3}
\end{align}
where the asterisk denotes complex conjugation. In FFE, the wave amplitude is defined as 
$$a_{\bfm{k}_i} = \frac{\delta E(\bfm{k}_i)}{\sqrt{2 \pi \omega_i}},$$ 
where $|\delta E(\bfm{k}_i)|$ is the amplitude of the electric (or magnetic) field with wavevector $\bfm{k}_i$ and angular frequency $\omega_i$. The phase of $\delta E(\bfm{k}_i)$ evolves on a fast timescale, set by the wave angular frequency, and on a slow timescale due to the resonant interactions.
 
It can be shown from Equations~(\ref{eq:three_wave_rescaled_1})--(\ref{eq:three_wave_rescaled_3}) that the quantities $\mathcal{F} \equiv |a_{\bfm{k}_1}|^2 + |a_{\bfm{k}_2}|^2$, $\mathcal{G} \equiv |a_{\bfm{k}_2}|^2 - |a_{\bfm{k}_3}|^2,$ and $\mathcal{H} \equiv i V (a_{\bfm{k}_1} a_{\bfm{k}_2}^* a_{\bfm{k}_3}^* + a_{\bfm{k}_1}^* a_{\bfm{k}_2} a_{\bfm{k}_3})$ are conserved. The conservation equations for $\mathcal{F}$ and $\mathcal{G}$ are called the Manley-Rowe relations, while $\mathcal{H}$ is the Hamiltonian of the system with canonical coordinates $q_i = a_{\bfm{k}_i}$ and $p_i = a_{\bfm{k}_i}^*$. The terms $|a_{\bfm{k}_i}|^2$ can be interpreted as the number of mode quanta (or wave occupation number), $n_{\bfm{k}_i}$, with the total energy contained in mode $i$ given by $U_{\bfm{k}_i} = n_{\bfm{k}_i} \omega_{\bfm{k}_i} = \omega_{\bfm{k}_i} |a_{\bfm{k}_i}|^2.$ The independent variables in the three-wave equations~(\ref{eq:three_wave_rescaled_1})--(\ref{eq:three_wave_rescaled_3}) are $|a_{\bfm{k}_i}|$ and $\Delta \phi \equiv \phi_{\bfm{k}_1} - \phi_{\bfm{k}_2} - \phi_{\bfm{k}_3},$ where $\phi_{\bfm{k}_i}$ are the wave phases. That is, the evolution equations for the individual phases, $\dot{\phi}_{\bfm{k}_i}$, can be determined from the governing equations for $|a_{\bfm{k}_i}|$ and $\Delta \phi$. Given the three constants of motion $\mathcal{F}, \mathcal{G}, \text{ and } \mathcal{H},$ the evolution equations are completely integrable.

In the limit where $|a_{\bfm{k}_1}| \gg |a_{\bfm{k}_2}|$ and $|a_{\bfm{k}_3}|$, the system is susceptible to the parametric decay instability, in which mode $1$ exponentially decays and transfers its energy to modes $2$ and $3$. We define the approximate interaction timescale for this process as the `e-folding' time for the growth of modes $2$ and $3$ in units of the periods of wave $1$:
\begin{equation}
    \frac{t_{\rm int}^{(1)}}{T_{\rm w,1}} \sim \frac{\omega_1}{2\pi}(V |{a}_{\bfm{k_1}}|)^{-1} \sim \frac{1}{\pi} \cdot \left(\frac{\sin^2 (\theta) \delta B_{\bfm{k_1}}}{B_0}\right)^{-1}.
    \label{eq:first_order_timescale}
\end{equation}
In the last step, we assumed that the interaction coefficient scales according to Equation~(\ref{eq:vfaa_limit}).

In this limit, $t_{\rm int}^{(1)}$ is \textit{independent} of $|a_{\bfm{k}_2}|$, $|a_{\bfm{k}_3}|$, and $\Delta \phi$. Assuming a narrow spectrum of propagating waves, the parametric decay of FMS modes is the dominant process during their escape from the inner magnetosphere of a magnetar. This process becomes increasingly important as the FMS waves propagate away from the magnetar and $\delta B_{\bfm{k}_1} / B_0$ grows.

\subsection{Evolution of a Spectrum of Resonant Modes} \label{subsec:Evolution of a Spectrum of Resonant Modes}
The discussion in the previous section applies to a system composed of only three resonantly interacting waves. For the more general case of a broad spectrum of interacting modes, each mode can participate in multiple resonantly interacting triads. In this case, the three-wave equations must incorporate all possible interactions by expressing Equations~(\ref{eq:three_wave_rescaled_1})--(\ref{eq:three_wave_rescaled_3}) as integrals, which we do in this section. Following \cite{golbraikh_2023ApJ...957..102G}, we do not consider the $A \leftrightarrow A + A$ process in the following equations. We first define $\tilde{V}^{F}_{123}$ and $\tilde{V}^{A}_{123}$ as:
\begin{align}
    \tilde{V}^{F}_{123} \equiv \quad &V_{FFA}(\bfm{k_1}, \bfm{k_2}, \bfm{k_3})  \delta (\bfm{k_1 - k_2 - k_3})   \\
    \tilde{V}^{A}_{123} \equiv \quad &V_{FAA}(\bfm{k_1}, \bfm{k_2}, \bfm{k_3})  \delta (\bfm{k_1 - k_2 - k_3}).
\end{align}
Here, $\tilde{V}^{F}_{123}$ and $\tilde{V}^{A}_{123} $ are the interaction coefficients, which also include the wavevector resonance condition.  

Let $\tilde{A}_{\bfm{k}_i}$ and $\tilde{a}_{\bfm{k}_i}$ denote the complex Fourier amplitudes for FMS and Alfvén waves, respectively. These are related to the Fourier amplitudes of the wave electric fields as $\tilde{A}_{\bfm{k}_i} \equiv \delta \hat{E}_{\bfm{k}_i}^F / (2 \pi \omega^F_{\bfm{k}_i})^{1/2}$ for FMS waves and $\tilde{a}_{\bfm{k}_i} \equiv \delta \hat{E}_{\bfm{k}_i}^A / (2 \pi \omega^A_{\bfm{k}_i})^{1/2}$ for Alfvén waves.\footnote{Note that $\tilde{A}_{\bfm{k}_i}$ and $\tilde{a}_{\bfm{k}_i}$ have different units than the $a_{\bfm{k}_i}$ in Equations~(\ref{eq:three_wave_rescaled_1})--(\ref{eq:three_wave_rescaled_3}) since the former contain the Fourier amplitudes and their evolution equations are integrals in $k$-space.} Then, the evolution equations for the Fourier amplitudes modes are given by:
\begin{align}
    \frac{{\rm d} \tilde{A}_{\bfm{k_1}}}{{\rm d}t} &=  -i \omega^F_{\bfm{k_1}} \tilde{A}_{\bfm{k_1}} +  i \iint \left(\tilde{V}^{F}_{123} \tilde{A}_{\bfm{k_2}} \tilde{a}_{\bfm{k_3}}+\right. \nonumber \\ & \left.   \tilde{V}^{F}_{213} \tilde{A}_{\bfm{k_2}} \tilde{a}_{\bfm{k_3}}^* + \frac{1}{2} \tilde{V}^{A}_{123} \tilde{a}_{\bfm{k_2}} \tilde{a}_{\bfm{k_3}} \right) {\rm d}^3\bfm{k_2} {\rm d}^3 \bfm{k_3} \label{eq:Adot} \\
    \frac{{\rm d} \tilde{a}_{\bfm{k_1}}}{{\rm d}t} &=  -i \omega^A_{\bfm{k_1}} \tilde{a}_{\bfm{k_1}} +i \iint \left(\tilde{V}^{F}_{231} \tilde{A}_{\bfm{k_2}} \tilde{A}_{\bfm{k_3}}^* + \right.\nonumber 
    \\ & \left. \quad \tilde{V}^{A}_{231} \tilde{A}_{\bfm{k_2}} \tilde{a}_{\bfm{k_3}}^*  \right) {\rm d}^3\bfm{k_2} 
{\rm d}^3 \bfm{k_3} \label{eq:adot}.
\end{align}
The phases of $\tilde a_{\bfm{k}_i}$ ($\tilde A_{\bfm{k}_i}$) in Equations~(\ref{eq:Adot}) and~(\ref{eq:adot}) rapidly vary on the angular frequency of waves as $\tilde a_{\bfm{k}_i} = |\tilde a_{\bfm{k}_i}| \exp[i(\omega (\bfm{k}_i) t + \phi({\bfm{k}_i})]$. The slow variation of the wave amplitudes can be obtained by time-averaging the equations over the fast phase variations. However, if the spectrum of waves is broad and the waves are randomly phased, the time-averaged integrals on the RHS of Equations~(\ref{eq:Adot}) and~(\ref{eq:adot}) vanish (see Appendix~\ref{appendix:many_wave_equation}). This implies that $|\tilde a_{\bfm{k}_i}|$ do not change at the lowest order of nonlinear interactions, i.e., on the timescale $t^{(1)}_{\rm int}$ given by Equation~(\ref{eq:first_order_timescale}). To reveal the secular nonlinear interaction timescale, the RHS of Equations~(\ref{eq:Adot}) and~(\ref{eq:adot}) must be expanded to higher order by taking the time derivative of the respective equations. The wave amplitudes evolve according to the following equations:
\begin{align}
    \frac{\dd N_{\bfm{k_1}}}{\dd t} &= \iint \left[-W^{F}_{123} \left(N_{\bfm{k_1}} ( n_{\bfm{k_3}} + N_{\bfm{k_2}}) - N_{\bfm{k_2}} n_{\bfm{k_3}} \right) + \right.  \nonumber \\ & \left. W^{F}_{213} \left( N_{\bfm{k_2}} ( n_{\bfm{k_3}} + N_{\bfm{k_1}}) - N_{\bfm{k_1}} n_{\bfm{k_3}}  \right) + \right. \nonumber \\ & \left. \frac{1}{2}  W^{A}_{123} \left( n_{\bfm{k_2}} n_{\bfm{k_3}} - N_{\bfm{k_1}} (n_{\bfm{k_3}} + n_{\bfm{k_2}})  \right)
    \right] \dd^3 \bfm{k_2} \dd^3\bfm{k_3}  
    \label{eq:Ndot} \\
    \frac{\dd  n_{\bfm{k_1}}}{\dd t} &= \iint \left[
    W^{F}_{231} \left(N_{\bfm{k_1}} ( n_{\bfm{k_2}} + N_{\bfm{k_3}}) - N_{\bfm{k_3}} n_{\bfm{k_2}}\right) \right. \nonumber \\ &
    \left. - W^{A}_{231} \left(n_{\bfm{k_3}} n_{\bfm{k_1}} - N_{\bfm{k_2}} (n_{\bfm{k_1}} + n_{\bfm{k_3}}) \right)
    \right] \dd^3 \bfm{k_2} \dd^3\bfm{k_3}  \label{eq:ndot},
\end{align}
where $n_{\bfm{k}_i}$ ($N_{\bfm{k}_i}$) are the mode quanta, $W^{F}_{ijk} = 2 \pi |\tilde{V}^F_{ijk}|^2 \delta(\omega^F_{\bfm{k}_j} + \omega^{A}_{\bfm{k}_k} - \omega^F_{\bfm{k}_i})$, and $W^{A}_{ijk} = 2 \pi |\tilde{V}^A_{ijk}|^2 \delta(\omega^A_{\bfm{k}_j} + \omega^{A}_{\bfm{k}_k} - \omega^F_{\bfm{k}_i})$. The mode quanta are given by $n_{\bfm{k}} \delta(\bfm{k} - \bfm{k}_i) = \langle \tilde a_{\bfm{k}} \tilde a^*_{\bfm{k}_i} \rangle$, where $\langle \cdots \rangle$ denotes time averaging over many wave periods.

The characteristic timescale $t_{\rm int}^{(2)}$ for this interaction is given by \citep[see][]{Lyubarsky_crab_2019MNRAS.483.1731L}:

\begin{equation}
    \frac{t_{\rm int}^{(2)}}{T_{\rm w}} \sim \left(\frac{\delta B_{\bfm{k_i}}}{B_0}\right)^{-2},
    \label{eq:second_order_timescale}
\end{equation}
The above estimate ignores the geometric dependence of the wavevectors in the interaction coefficients.

In the absence of phase averaging, when the spectrum of energetic FMS waves is narrow, $\dot{N}_{\bfm{k}}$ and $\dot{n}_{\bfm{k}}$ are nonzero even at lowest order. In the early stages of evolution, such a system resembles a series of uncoupled three-wave interactions, where each triad contains an energetic FMS wave and two low-amplitude waves. For narrowband FMS signals produced in the inner magnetar magnetosphere, with $\Delta \nu \ll \nu$, the nonlinear evolution occurs on the timescale $t^{(1)}_{\rm int}$. For relatively broad-spectrum FMS signals, with $\Delta \nu \gtrsim \nu$, the evolution proceeds on the timescale $t^{(2)}_{\rm int}$. In this paper, we investigate both cases via FFE simulations and demonstrate that a significant fraction of the FMS energy is converted into resonant Alfvén waves occupying a large $k_\perp$ region of $k$-space.

\subsection{Momentum Conservation} \label{subsec:Momentum Consevation}
In the preceding sections, we discussed how FMS waves can decay into other FMS waves and Alfvén waves via resonant interactions. An important question is whether these product linear modes conserve the Poynting flux (momentum) of the original FMS wave. To address this, we consider the $F \leftrightarrow A + A$ process: an FMS wave with wavevector $\bfm{k}_1$ transforms into two Alfvén waves with wavevectors $\bfm{k}_2$ and $\bfm{k}_3$. Let $\bfm{S}_1$, $\bfm{S}_2$, and $\bfm{S}_3$ denote the corresponding Poynting fluxes of the three waves,
\begin{align}
    \bfm{S}_1 &= (c / 4\pi) \left\langle\delta E(k_1)\right\rangle^2 \bfm{\hat{k}_1} \, \, \, \, = n_{\bfm{k_1}} \omega_1 \, c \, \bfm{\hat{k}_1} \\
    \bfm{S}_2 &= (c / 4\pi) \left\langle\delta E(k_2)\right\rangle^2 \bfm{\hat{k}_{2,z}}  = n_{\bfm{k_2}} \omega_2 \, c \, \mathrm{sign}(k_{2, z}) \,  \bfm{\hat{z}} \\
    \bfm{S}_3 &= (c / 4\pi) \left\langle\delta E(k_3)\right\rangle^2 \bfm{\hat{k}_{3,z}}  = n_{\bfm{k_3}} \omega_3 \, c \, \mathrm{sign}(k_{3, z}) \, \bfm{\hat{z}},
    \label{eq:poynting_flux_three_waves}
\end{align}
where $\delta E(k_i)$ is the electric field amplitude of mode $i$ and  $\left\langle\cdots \right\rangle$ denotes the RMS value over a wave period. Using the conservation of $\mathcal{F}$ and $\mathcal{G}$, it follows that $n_{\bfm{k_2}} = \mathcal{F} - n_{\bfm{k_1}}$ and $n_{\bfm{k_3}} = \mathcal{F} - \mathcal{G} - n_{\bfm{k_1}}$. Then, the total Poynting flux in the $z$-direction is:
\begin{align}
    n_{\bfm{k_1}} k_{1,z}  + n_{\bfm{k_2}} k_{2,z} + n_{\bfm{k_3}} k_{3,z} = \notag \\
    \mathcal{F}(k_{2,z} + k_{3,z}) - \mathcal{G}k_{3,z} + n_{\bfm{k_1}} & (k_{1,z} - k_{2,z} - k_{3,z}) 
\end{align}
Equation~(\ref{eq:resonance_k}) implies that the RHS of the above equation is constant, and the three-wave process conserves Poynting flux in $z$:
\begin{equation}
\begin{aligned}
    &\frac{\dd }{\dd t} (n_{\bfm{k_1}} k_{1,z}  + n_{\bfm{k_2}} k_{2,z} + n_{\bfm{k_3}} k_{3,z}) = 0, \\
    &\implies c^2 \frac{\dd }{\dd t} (n_{\bfm{k_1}}  k_{1,z}  + n_{\bfm{k_2}} k_{2,z} + n_{\bfm{k_3}} k_{3,z})\bfm{\hat{z}} = \bfm{0},\\
    &\implies c \frac{\dd }{\dd t} (n_{\bfm{k_1}} \omega_1 (\bfm{\hat{k}_1} \cdot \bfm{\hat{z}}) + 
    n_{\bfm{k_2}} \omega_2  \, \mathrm{sign}(k_{2, z}) + \\
    & \quad \qquad n_{\bfm{k_3}} \omega_3  \, \mathrm{sign}(k_{3, z})) \bfm{\hat{z}} = \bfm{0}, \notag \\
    &\implies  \bfm{\dot{S}_1} \cdot \bfm{\hat{z}} = -(\bfm{\dot{S}_2}  + \bfm{\dot{S}_3}).
\end{aligned}
\label{eq:conservation_of_S}
\end{equation}
Thus, the two Alfvén waves carry the $z$-component of the Poynting flux of the decaying FMS wave. However, if $\bfm{k}_1$ has a component perpendicular to the background field, the two Alfvén waves cannot carry $\bfm{S}_{1,\perp}$ since $\bfm{S}_{2,\perp} = \bfm{S}_{3,\perp} = \bfm{0}$.

Momentum conservation still holds in FFE, and the `missing' Poynting flux resides in higher-order fields that are not linear modes of the system. Specifically, $\bfm{S}_{1,\perp}$ is accounted for by the Poynting flux produced by a spatially and temporally uniform electric field $\bfm{\bar{E}^{(2)}}$ with $\omega = 0$ and $\bfm{k} = \bfm{0}$. The corresponding Poynting flux from this nonlinear electric field is given by $\bfm{S}_{\rm NL} = (c / 4\pi) \, \bfm{\bar{E}^{(2)}} \times B_0 \, \bfm{\hat{z}}$. In Appendix~\ref{appendix:momentum conservation}, we show that
\begin{equation}
    \bfm{S}_{1, \perp}(t) + \bfm{S}_{\rm NL}(t) = \bfm{S}_{1, \perp}(t=0) = \mathrm{const.}
\end{equation}
In summary, when FMS waves decay into two Alfvén waves, they also produce nonlinear electromagnetic fields carrying Poynting flux along $\bfm{\hat{k}}_{1,\perp}$, ensuring momentum conservation. A similar process occurs for initially broad spectra of FMS waves. Fully nonlinear FFE simulations show that when the FMS waves deposit most of their energy into Alfvén waves, the perpendicular momentum of the original FMS waves is carried solely by the nonlinear fields, $\bfm{S}_{\rm NL}$.

\begin{figure*}
    \centering
    \includegraphics[width=0.99 \textwidth]{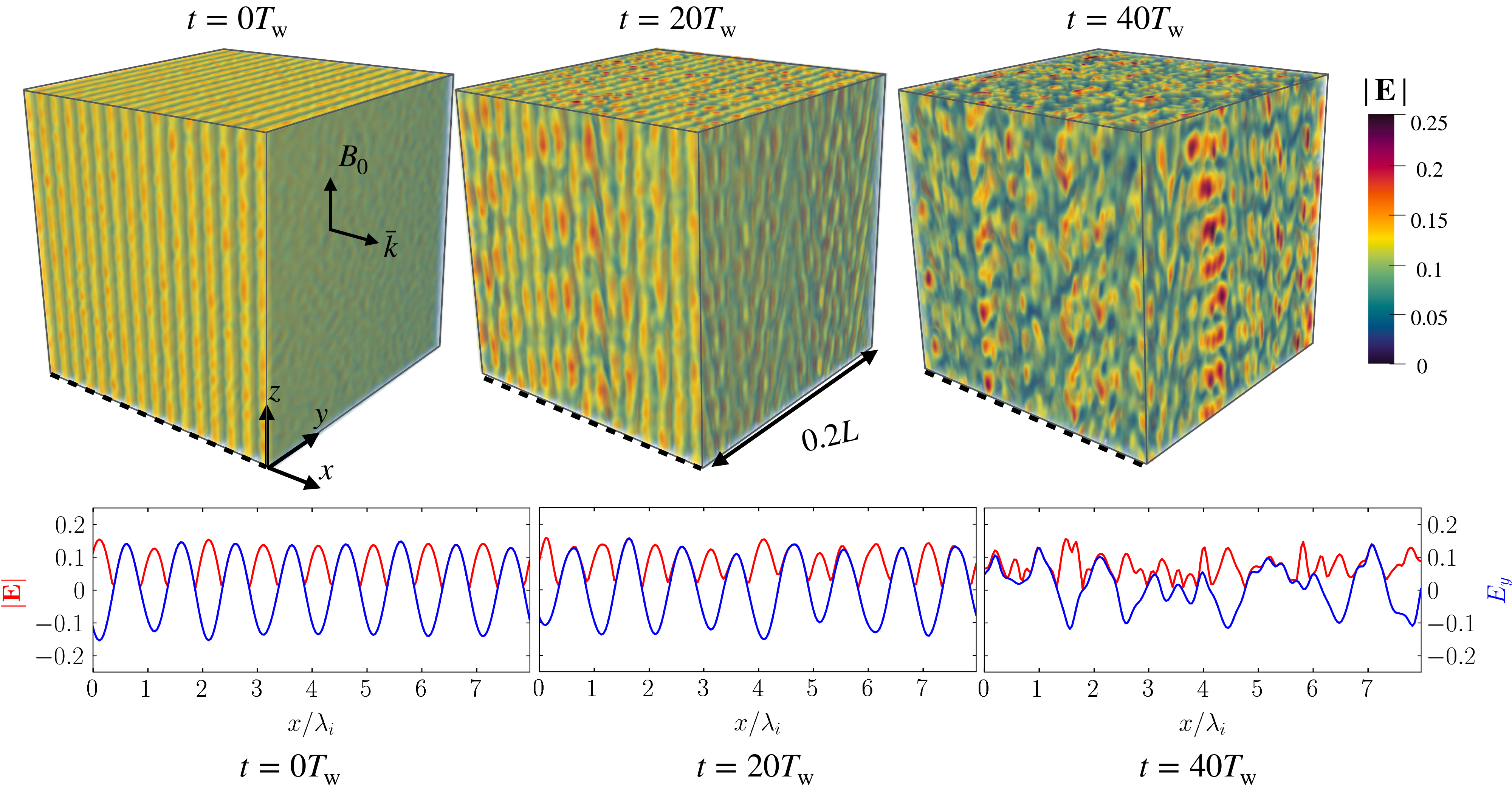}
\caption{Interaction dynamics of FMS waves with a tenuous Alfvén wave background (``\texttt{Mono-90}''). \textbf{Top:} Volume rendering of the electric field magnitude in a subset of the simulation domain. The three panels show the initial state (left) and evolved states at 20 $T_{\rm w}$ (middle) and 40 $T_{\rm w}$ (right). The wavevector of the injected FMS mode is along the $x$-axis, and the background field is along the $z$-axis. Over time, FMS waves transfer energy to a broad spectrum of FMS and Alfvén waves via resonant three-wave interactions. \textbf{Bottom:} One-dimensional slices along the $x$-axis of the electric field magnitude (red) and the $y$-component of the electric field (blue), taken along the black dashed lines in the top panels. By 40 $T_{\rm w}$, the initial FMS wave has significantly evolved, exhibiting strong distortion due to nonlinear interactions.}
    \label{fig:electric_field_mag}
\end{figure*}

\section{Simulations} \label{section:Simulations}

We model nonlinear wave interactions using a high-order finite-volume FFE method \citep{Jens_code_1_2021A&A...647A..57M,Jens_code_2_2021A&A...647A..58M}, adopting optimizations for weak Alfvénic turbulence models performed with the same code \citep{alfven_turbulence_ripperda_2021JPlPh..87e9012R}. Our numerical framework uses the infrastructure of the \textsc{Einstein Toolkit}\footnote{\url{http://einsteintoolkit.org}} \citep{Loeffler2012} and benefits from the \textsc{Carpet} driver \citep{Goodale2002a,Schnetter2004}. All simulations in this work use Cartesian coordinates $x,y,z \in [0,L] \times [0,L] \times [0,L]$ with uniform resolutions of $N_{xyz} \times N_{xyz} \times N_{xyz} = 1024^3$ cells and a cube size of $L=1$. We use seventh-order spatial reconstruction \citep[MP7,][]{SURESH199783} with a CFL factor of $f_{\rm CFL} = 0.25$ and hyperbolic/parabolic cleaning of divergence errors. We integrate the evolution of the electric and magnetic fields in time with the corresponding force-free current, $\mathbf{j}$, and charge density, $\rho$, while enforcing the constraints
\begin{align}
\rho\mathbf{E}+\mathbf{j}/c\times\mathbf{B}=\mathbf{0}&\qquad\text{(force-free condition)},\label{eq:ffbalance}\\
\mathbf{E}\cdot\mathbf{B}=0&\qquad\text{(ideal fields)}\label{eq:FFI},\\
    E<B&\qquad\text{(magnetic dominance)}\label{eq:FFII}.
\end{align}
We study the nonlinear wave interactions between FMS and Alfvén waves using two sets of simulations. In the first set, FMS waves are initialized to be energetically dominant over AWs (Section~\ref{sec:FWSims}). In the second set, FMS and Alfvén waves initially have equal energy (Section~\ref{subsec:FMS-Alfven Equipartition}). All models are summarized in Table~\ref{tab:sims}.

\subsection{Initial conditions} \label{subsec:General Setup}

Our simulations initialize a background magnetic field $\bfm{B}_0 \equiv B_0 \bfm{\hat{z}}$, on which we superimpose different wave modes with amplitudes $0 \leq \xi_{\bfm{k}} < 1$ and random phases $\phi_{\bfm{k}} \in [0, 2\pi]$. The total EM fields in the simulation are $\bfm{E} = \delta \bfm{E} = \sum_{i=A,F} \sum_{\bfm{k}} \xi_i(\bfm{k}) \cos(\bfm{k \cdot r} + \phi_i) \bfm{\hat{e}}_{i}$ and $\bfm{B} = \bfm{B}_0 + \delta \bfm{B} = B_0 \bfm{\hat{z}} + \sum_{i=A,F} \sum_{\bfm{k}} \xi_i(\bfm{k}) \cos(\bfm{k \cdot r} + \phi_i) \bfm{\hat{b}}_i$, where $\bfm{\hat{e}}_i$ and $\bfm{\hat{b}}_i$ are the polarization vectors for the electric and magnetic fields of FMS and Alfvén waves, respectively (see Appendix~\ref{appendix:three_wave_interactions}).\footnote{Since the sum of linear modes is not necessarily force-free, i.e., does not exactly satisfy $\bfm{E \cdot B} = 0$ at higher order in wave amplitudes, we enforce condition~(\ref{eq:FFI}) before the first timestep. This correction introduces low-amplitude, transient structures that have no dynamical importance at later times.}

In the first set of simulations (Section~\ref{sec:FWSims}), we initialize the FMS and AWs with two different distributions of wavevectors: FMS waves start with a Gaussian spectrum and AWs with a flat spectrum, where each wavevector $\bfm{k}$ carries the same energy. Wave amplitudes for the Gaussian spectra are computed as $\xi(\bfm{k}) = A_0 \exp(-|\bfm{k - \bar{\bfm{k}}}|^2 / (2\sigma_k^2))$, where $A_0$ is a normalization constant, $\bar{\bfm{k}}$ is the mean wavevector, and $\sigma_k$ is the width of the spectrum. $A_0$ is set such that the total FMS wave energy matches a target value at the start of the simulation. We ensure that $U_{\rm w} / U_{\rm bg} \lesssim 1$, where $U_{\rm w}$ is the total wave energy summing all modes, $U_{\rm w} = \sum_{i=A,F} \sum_{\bfm{k}} \xi_i^2(\bfm{k}) / 2$, and $U_{\rm bg} = B_0^2 / 2$ is the energy of the background magnetic field.

We perform an additional simulation (Section~\ref{subsec:FMS-Alfven Equipartition}) in which the FMS and AW energies are initialized in equipartition. The initial FMS spectrum is Gaussian, centered at $\bar{\bfm{k}}_1 = (\bar{\bfm{k}}_{1,\perp}, \bar{k}_{1,z})$, with a peak frequency $\omega_1 = |\bar{\bfm{k}}_1|$. The AW spectrum is initialized with two Gaussians centered at $\bar{\bfm{k}}_2 = (\bar{\bfm{k}}_{1,\perp}, \frac{1}{2}(|\bar{\bfm{k}}_1| + |\bar{k}_{1,z}|))$ and $\bar{\bfm{k}}_3 = (\bar{\bfm{k}}_{1,\perp}, \frac{1}{2}(|\bar{\bfm{k}}_1| - |\bar{k}_{1,z}|))$, chosen such that their frequencies satisfy the resonance condition $\omega_1 = \omega_2 + \omega_3$.

\subsection{Scale separation requirements} \label{subsec:scale_sep}

On a numerical grid with finite resolution, the minimally expected frequency mismatch in the wave triads is of the order $\Delta \omega \sim 2 \pi c / L$. The waves thus lose resonance on a timescale $t_{\rm dec} \sim 2 \pi / \Delta \omega \lesssim L / c$. We identify the requirements for mildly nonlinear FMS waves with a wavelength $\lambda_i$ to efficiently participate in resonant three-wave interactions. First, the nonlinear interaction timescales, $t^{(1)}_{\rm int}$ and $t^{(2)}_{\rm int}$, must be shorter than $t_{\rm dec}$. For narrow spectra, Equation~(\ref{eq:first_order_timescale}) demands $\delta B_{\bfm{k_i}} / B_0 \gtrsim \lambda_i / L$; for broad spectra, Equation~(\ref{eq:second_order_timescale}) requires $\delta B_{\bfm{k_i}} / B_0 \gtrsim (\lambda_i / L)^{1/2}$. Studying mildly nonlinear waves therefore requires a significant separation between the box size and the FMS wavelength. Second, the FMS wavelength and grid scale must be well separated to allow for a large range of $k$ of excited waves, such that $\lambda_i \gg \Delta_{xyz}$, where $\Delta_{xyz} = L / N_{xyz}$ is the grid cell size. The initial simulation conditions satisfy the scale requirement $d_{xyz} < \lambda_i < L$, with $\lambda_i / L \sim 2 \times 10^{-2}$ and $\lambda_i / \Delta_{xyz} \sim 20$.

\begin{figure*}
    \centering
    \includegraphics[width=0.98\textwidth]{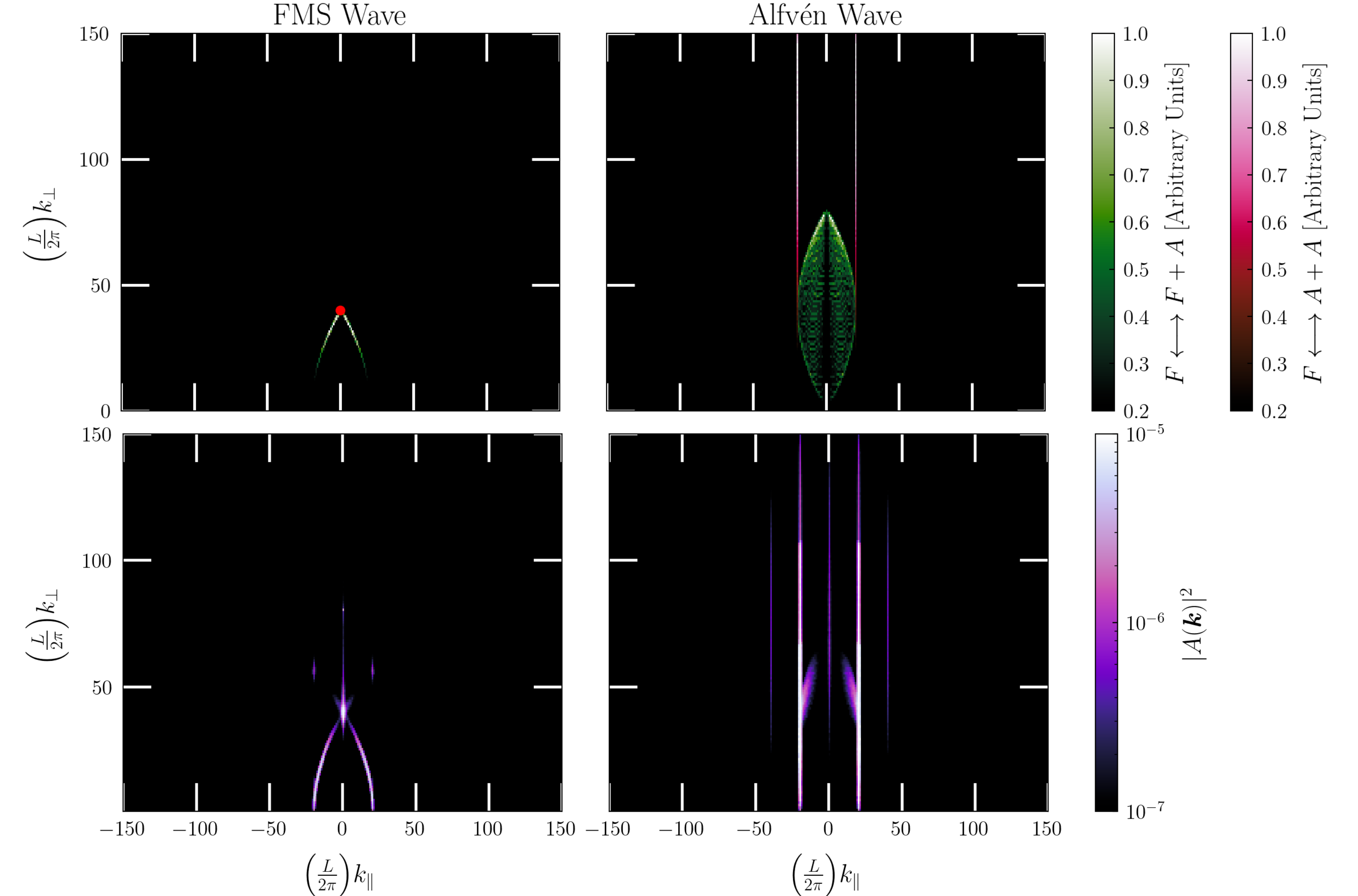}
    \caption{Comparison between theoretical expectations from parametric decay (top) and the spectrum of secondary waves excited in a numerical simulation (bottom). Top row: Theoretical $k_{\parallel}$-$k_{\perp}$ distribution of modes expected from the parametric decay of a single FMS wave (initial mode is indicated as a red dot). FMS waves and AWs produced from the $F \longleftrightarrow F + A$ process are colored green, while pairs of AWs produced from $F \longleftrightarrow A + A$ are shown in pink. Bottom row: Spectrum calculated in simulation ``\texttt{Mono-90}'' at $t = 30 T_w$. The simulation data closely resemble the theoretical expectation, indicating that three-wave interactions are a primary driver of the system's dynamics.}
    \label{fig:kpar_k_perp_spectrum}
\end{figure*}

\subsection{Dynamics of FMS waves interacting with a tenuous AW background}
\label{sec:FWSims}

In our first set of simulations, we initialize FMS waves with a Gaussian wave spectrum, and AWs with equal energy at each wavenumber $\bfm{k}$ for $k < k_{\rm cut} = 70 \times 2\pi / L$. In all simulations of this section, the total AW energy is much smaller than the total FMS wave energy. We then vary the spectral width, $\sigma_k$, the orientation of the FMS wavevector, $\bfm{\bar{k}}$, with respect to the background field, and the total initial FMS wave energy.

We illustrate the setup and wave evolution in Figure~\ref{fig:electric_field_mag}. The simulation ``\texttt{Mono-90}'' initializes a monochromatic FMS wave propagating along the $x$-axis, perpendicular to the background magnetic field. The panels show the time evolution of the electric field y-component, $E_y$, and its total magnitude, $|\mathbf{E}|$. The wave evolves considerably over $40 T_{\rm w}$, corresponding to one box light-crossing time. The initial structure of the EM fields is practically erased. The presence of intermittent, small-scale patches of high electric field strength suggests that a broad spectrum of waves is excited. As we show below, the initial FMS wave efficiently converts its energy into a mix of FMS and Alfvén waves.

\subsubsection{Resonant excitation of secondary waves}

We now examine how the primary FMS waves transfer their energy to secondary FMS and Alfvén waves via resonant interactions. To do so, we compute the $k_{\perp}$-$k_{\parallel}$ spectra for both modes in the simulation ``\texttt{Mono-90}'' and compare them with theoretical expectations.

The theoretical spectra are constructed as follows. For the initial FMS wave with wavevector $\bfm{\bar{k}_1}$ and angular frequency $\omega_1 \equiv \bar{k}_1 c$, we find FMS-AW and AW-AW pairs with wavevectors and angular frequencies $(\bfm{k}_2, \bfm{k}_3 ; \omega_2, \omega_3)$ that satisfy the resonant conditions (\ref{eq:resonance_k})-(\ref{eq:resonance_omega}) within a tolerance set by the grid spacing, $\Delta k$, where $\Delta k \equiv \Delta \omega / c = 2\pi / L$. We scan over each Cartesian component of $\bfm{k}_2$ and $\bfm{k}_3$ in the range $(-k_{\rm max}, k_{\rm max})$ in steps of $\Delta k$, where $2\pi / k_{\rm max}$ corresponds to the dissipative scale of the simulation (a few grid cells). The wave pairs satisfying the resonance are weighted by $w = e^V$, where $V$ is the respective interaction rate $V_{FFA}$ or $V_{FAA}$ (see Equations \ref{eq:vffa} and \ref{eq:vfaa}). The weighting captures the mode growth in the initial stages of the parametric decay. We then bin resonant modes in $k_{\parallel}$-$k_{\perp}$ space with respect to the background magnetic field, $k_{\parallel} = k_z$ and $k_{\perp} = (k_x^2 + k_y^2)^{1/2}$.

We synthesize all possible resonant pairs and their respective interaction rates in the top row of Figure~\ref{fig:kpar_k_perp_spectrum}. The red dot represents the initial FMS wave, and green and pink dots show resonant waves excited through $F \longleftrightarrow F + A$ and $F \longleftrightarrow A + A$ interactions, respectively. The $F \longleftrightarrow F + A$ process generates a complex distribution of resonant secondary FMS and Alfvén waves in the $k_{\parallel}$-$k_{\perp}$ space, but with a limited spectral extent. The secondary AWs have $|k_{\parallel}| \in (0, |k|)$ and $k_{\perp} \in (0, 2 |\bar{k}|)$, following from the resonance conditions. In contrast, the $F \longleftrightarrow A + A$ process creates AW pairs with only two possible frequencies (or equivalently, $k_{\parallel}$), but with a large range of $k_{\perp}$. The two frequencies are $\omega_{2,3} = (\bar{k}_1 \pm \bar{k}_{1, \parallel}) c / 2$ (see Sect.~\ref{subsection: Interactions of Three Resonant Modes}), appearing as two vertical pink stripes in the top right panel of Figure~\ref{fig:kpar_k_perp_spectrum}.

We compare our theoretical expectations with a simulation spectrum computed at $t / T_{\rm w} = 30$ in the bottom row of Figure~\ref{fig:kpar_k_perp_spectrum}. The 2D spectrum, $|A(\bfm{k})|^2$, of excited FMS and Alfvén waves is computed by projecting Fourier transforms of electric and magnetic fields along the FMS and Alfvén wave eigenvectors (see Equations A5–A8 in Appendix~\ref{appendix:three_wave_interactions}). The simulation spectrum reproduces key features of the theoretical expectations. Additionally, the simulations produce FMS and Alfvén modes with frequencies above the original FMS mode (Figure~\ref{fig:kpar_k_perp_spectrum}, bottom left panel). These waves cannot be produced from the parametric decay of the original wave and result from interactions of the produced secondary waves.

\begin{figure*}
    \centering
    \includegraphics[width=0.98\textwidth]{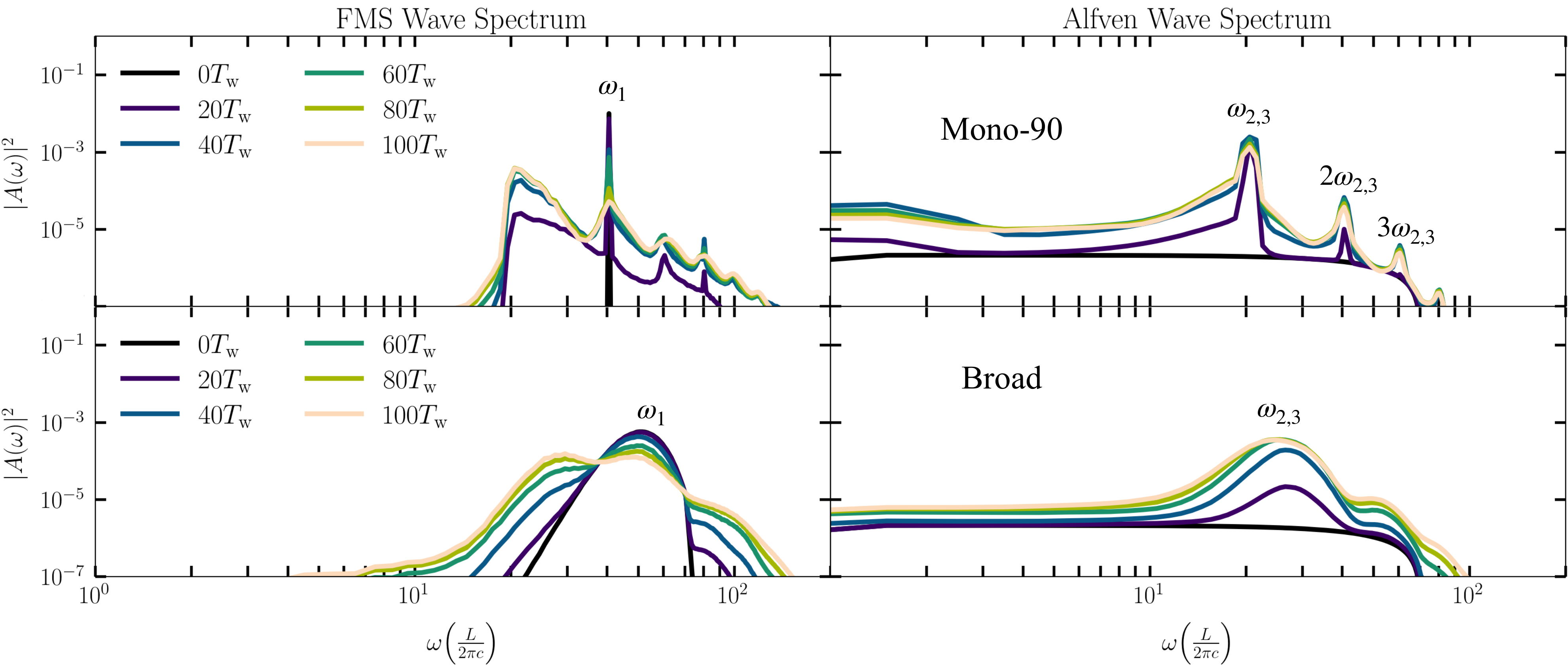}
    \caption{Frequency spectra of FMS and Alfvén waves for very narrow (``\texttt{Mono-90}'', top panels) and initially broad (``\texttt{Broad}'', bottom panels) FMS spectra, where $|A(\omega)|^2$ represents the total energy in each frequency bin. Time is scaled by $T_{\rm w}$, the period of FMS waves at the peak of the initial spectrum. Frequencies are normalized by $2 \pi L / c$, so that $\omega = 1$ corresponds to an FMS wave with wavelength equal to the simulation domain size. The initial FMS spectrum, centered around $\omega_1$, broadens significantly over time, reaching a saturated state with $\Delta \omega \gtrsim \omega$. Peaks labeled $\omega_{2,3}$ in the Alfvén wave spectrum arise from parametric decay ($F \longleftrightarrow A+A$, see Figure~\ref{fig:kpar_k_perp_spectrum}).
}
    \label{fig:fw_frequency_spectrum}
\end{figure*}

\subsubsection{Spectral evolution from the decay of narrow and broad-band FMS waves}
We now analyze the time evolution in frequency and $k_{\perp}$-space of two representative simulations, with an initially broad (``\texttt{Broad}'') and an initially monochromatic (``\texttt{Mono-90}'') FMS spectrum. Figure~\ref{fig:fw_frequency_spectrum} shows mode energies binned by frequencies obtained from spatial Fourier analysis ($k$) and evaluating the linear dispersion relations, $\omega_{\rm FMS} = |k| c$ and $\omega_{\rm AW} = |k_{\parallel}| c$. The total energy in a given frequency bin is $|A(\omega)|^2 = \sum_{\bfm{k'}} |A(\bfm{k'})|^2 \delta(\omega - \omega_{A,F}(\bfm{k'}))$, with all bins having the same width.

Both simulations evolve toward a saturated state that has a broad FMS and AW spectrum, with $\Delta \omega \gtrsim \omega$. Most of the FMS wave energy ends up in lower frequency FMS waves than the initial spectrum, consistent with decreasing frequency in the $F \longleftrightarrow F + A$ process. Additionally, both simulations show prominent AW production at the theoretically predicted frequencies via $F \longleftrightarrow A + A$ decay of the initial FMS wave (peaks marked as $\omega_{2,3}$). The spectral evolution stops when resonant waves drop in amplitude and de-correlation due to finite numerical resolution becomes important; see Section~\ref{subsec:scale_sep} and Appendix~\ref{appendix:Non-resonant Three-wave Interactions}.

\begin{figure*}
    \centering
    \includegraphics[width=0.98\textwidth]{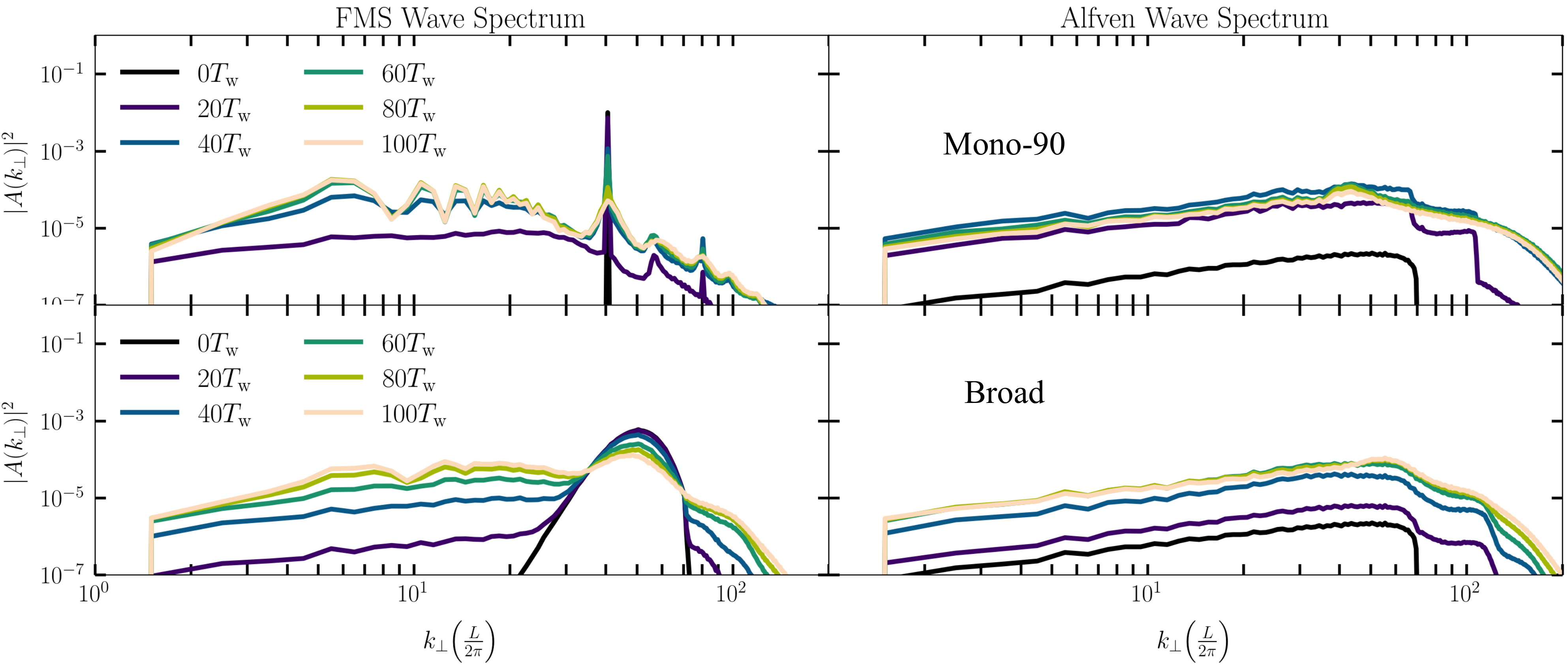}
    \caption{Same as Figure~\ref{fig:fw_frequency_spectrum}, but displaying the $k_\perp$ spectrum. Resonant interactions generate Alfvén waves with a broad $k_\perp$ distribution, due to the absence of $k_\perp$-dependence in the AW resonance condition (Equation~\ref{eq:resonance_omega}).}
    \label{fig:fw_kperp_spectrum}
\end{figure*}

Figure \ref{fig:fw_kperp_spectrum} shows corresponding spectra in $k_{\perp}$ space for the same two simulations (``\texttt{Broad}''/``\texttt{Mono-90}''). The total energy in a given wavenumber bin is $|A(k_{\perp})|^2 = \sum_{\bfm{k'}} |A(\bfm{k'})|^2 \delta(k_{\perp} - k_{\perp}' )$. FMS and Alfvén waves show very different $k_{\perp}$ spectra: AWs have considerably flatter $k_{\perp}$ spectra compared to FMS modes. This is because the wave-matching conditions in Equations~(\ref{eq:resonance_k}) and~(\ref{eq:resonance_omega}) do not constrain the $k_{\perp}$ values of resonant Alfvén waves, and FMS-decay into arbitrarily large $k_{\perp}$ waves is possible \citep{golbraikh_2023ApJ...957..102G}.

\subsubsection{Energy evolution for FMS and Alfvén waves with different initial FMS spectra}
\begin{figure*}
    \centering
    \includegraphics[width=0.85\textwidth]{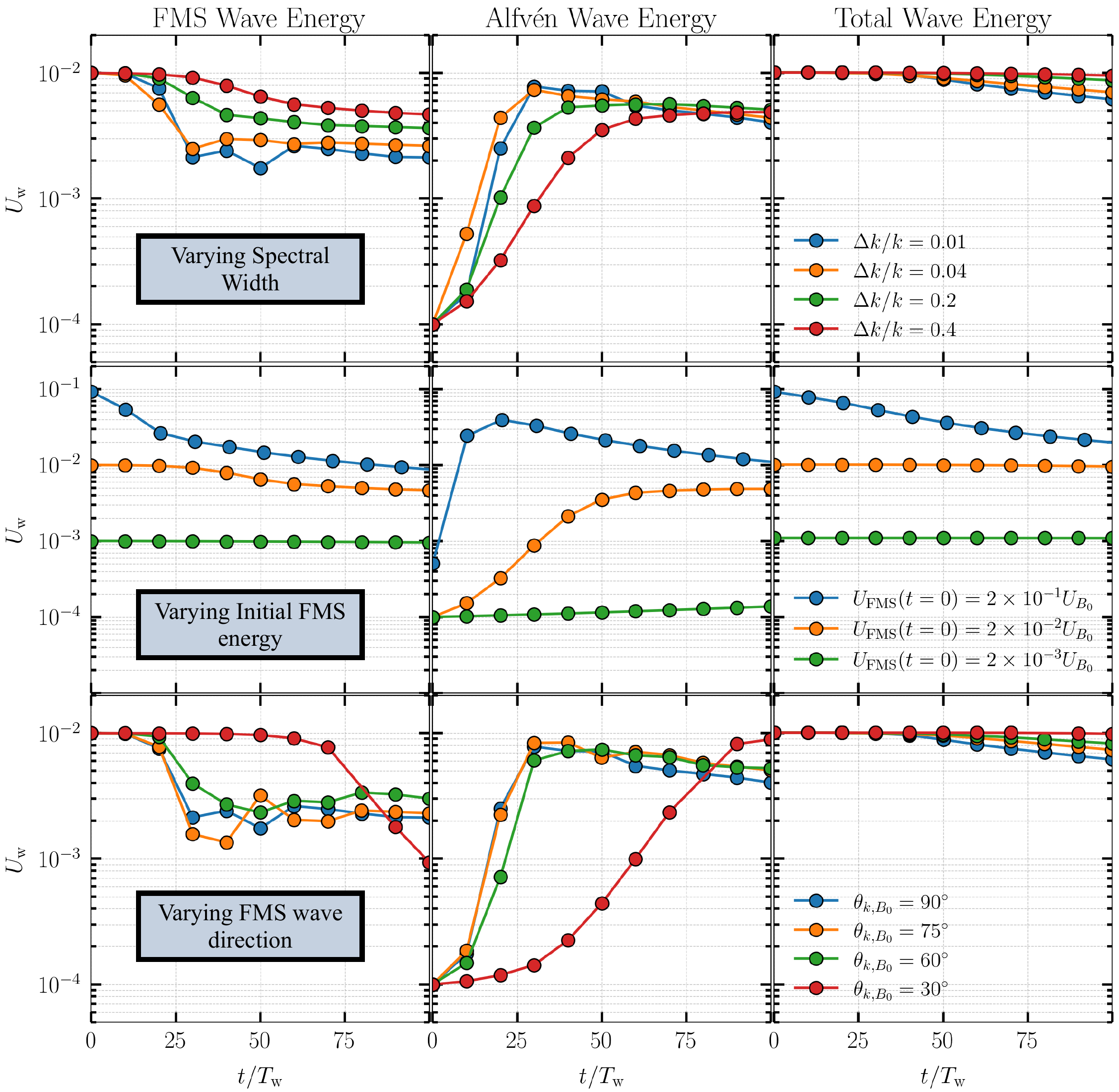}
    \caption{Time evolution of FMS, Alfvén, and total (FMS + Alfvén) wave energies for simulations in which the initial FMS energy exceeds the AW energy (Section~\ref{sec:FWSims}). In all cases, energy is transferred from FMS waves to Alfvén waves, with the conversion efficiency governed by three main factors: (i) a narrower initial FMS spectrum results in more efficient energy transfer, (ii) higher initial FMS energy drives stronger non-linear interactions, and (iii) oblique FMS propagation relative to the background magnetic field enhances mode coupling. In simulations with efficient FMS-to-AW conversion, the total wave energy decreases at late times due to excitation of AWs with large $k_{\perp}$ that dissipate at the grid scale. When the initial FMS energy is too low (e.g., the green curve in the middle row), resonance mismatches induced by the finite grid resolution suppress non-linear interactions.}
    \label{fig:fw_wave_energies}
\end{figure*}

In Figure~\ref{fig:fw_wave_energies}, we present the time evolution of FMS, Alfvén, and total wave energies for all simulations where FMS energy is initially greater than the AW energy. When non-linear interactions are well resolved, a substantial fraction of the initial FMS wave energy is efficiently transferred to Alfvén waves. 

The top row shows that FMS waves with narrower initial spectra decay into AWs more rapidly than those with broader spectra, consistent with the theoretical expectations discussed in Section~\ref{subsec:Evolution of a Spectrum of Resonant Modes}. For example, in the narrowband simulation shown in the top panel of Figure~\ref{fig:fw_wave_energies}, $\Delta k / k = 0.01$, and AWs reach approximate equipartition with FMS waves within $\sim 30$ wave periods. This timescale can be estimated using Equation~(\ref{eq:first_order_timescale}). For an initial FMS wave amplitude of $\delta B / B_0 = 0.1$ propagating perpendicularly to the background field, the $e$-folding time is $t^{(1)}/T_{\rm w} \sim 10 / \pi \approx 3.2$. Assuming the growth is dominated by the resonant triad with the largest coupling coefficient $V_{FFA/FAA}$, and that only a single wavevector $\bfm{k}$ dominates resonant interactions during early evolution, we can estimate the number of $e$-foldings needed to reach equipartition. The initial amplitude of the resonant AW is $\sim 1.7 \times 10^{-5}$. The number of $e$-foldings required to grow from this amplitude to that of the parent FMS wave is $\ln{(\delta B_{\rm FMS}(t=0) / \delta B_{\rm A}(t=0))} \approx 8.5$. This calculation implies a total interaction time of approximately $8.5 \times 3.2 \approx 27.2$ wave periods, which is in good agreement with our simulation.

The second row of Figure~\ref{fig:fw_wave_energies} demonstrates that the rate of non-linear interactions increases with the ratio of FMS-to-background field energies. However, when the initial FMS energy is too low, as in the $U_{\rm FMS} = 10^{-3}$ case in the second row, non-linear interactions become inefficient, and resonant energy transfer is suppressed due to finite grid resolution effects (see Section~\ref{subsec:scale_sep}). This case effectively sets a lower bound on the FMS wave energy that our simulations can reliably capture. Generally, whenever the interaction time of a resonant triad drops below its decorrelation timescale, the decay process stalls. For most simulations, this threshold occurs at $U_{\rm FMS} \sim 10^{-3}$, though the precise value depends on the 3D spectral distribution of the FMS waves.

Finally, the bottom row of Figure~\ref{fig:fw_wave_energies} shows results from simulations of monochromatic FMS waves propagating at different angles relative to the background magnetic field. We find that oblique FMS waves decay more rapidly than those propagating parallel to the field, indicating enhanced mode coupling at oblique angles. This happens due to larger non-linear currents produced by waves with a large $k_{\perp}$ component, resulting in more efficient growth rates of seed waves. Equation~(\ref{eq:vfaa_limit}) estimates that the interaction rate scales as $\sin^2 \theta$, where $\theta$ is the angle between the wavevector of the original FMS wave and the background field. 

At later times, due to the decrease in wave power and the decorrelation of resonant waves at finite resolution, the continuous transfer of FMS energy into Alfvén waves (AWs) ceases. In the following section, we analyze AWs with initial energies comparable to that of the FMS wave, allowing us to track the nonlinear interactions over longer times.

\subsection{Mode coupling in strong Alfvénic wave background} \label{subsec:FMS-Alfven Equipartition}

In this Section we consider FMS and Alfvén waves with initial spectra such that the total AW energy is equal to the FMS wave energy (`\texttt{F=A}'). The FMS spectrum is a Gaussian centered at $\bfm{\bar{k}_1} = (\bfm{\bar{k}_{1, \perp}}, \bar{k}_{1,z})$, while the AWs are initialized with two Gaussians centered at $\bfm{\bar{k}_2} = (\bfm{\bar{k}_{1,\perp}}, \frac{1}{2}(|\bar{k}_1| + |\bar{k}_{1,z}|))$ and $\bfm{\bar{k}_3} = (\bfm{\bar{k}_{1,\perp}},  \frac{1}{2}(|\bar{k}_1| - |\bar{k}_{1,z}|))$. The characteristic frequencies of the two Gaussians in the AW spectrum correspond to the frequencies expected from the parametric decay of FMS waves with wavevectors $\approx \bfm{k}_1$. 

\begin{figure*}
    \centering
    \includegraphics[width=0.98\textwidth]{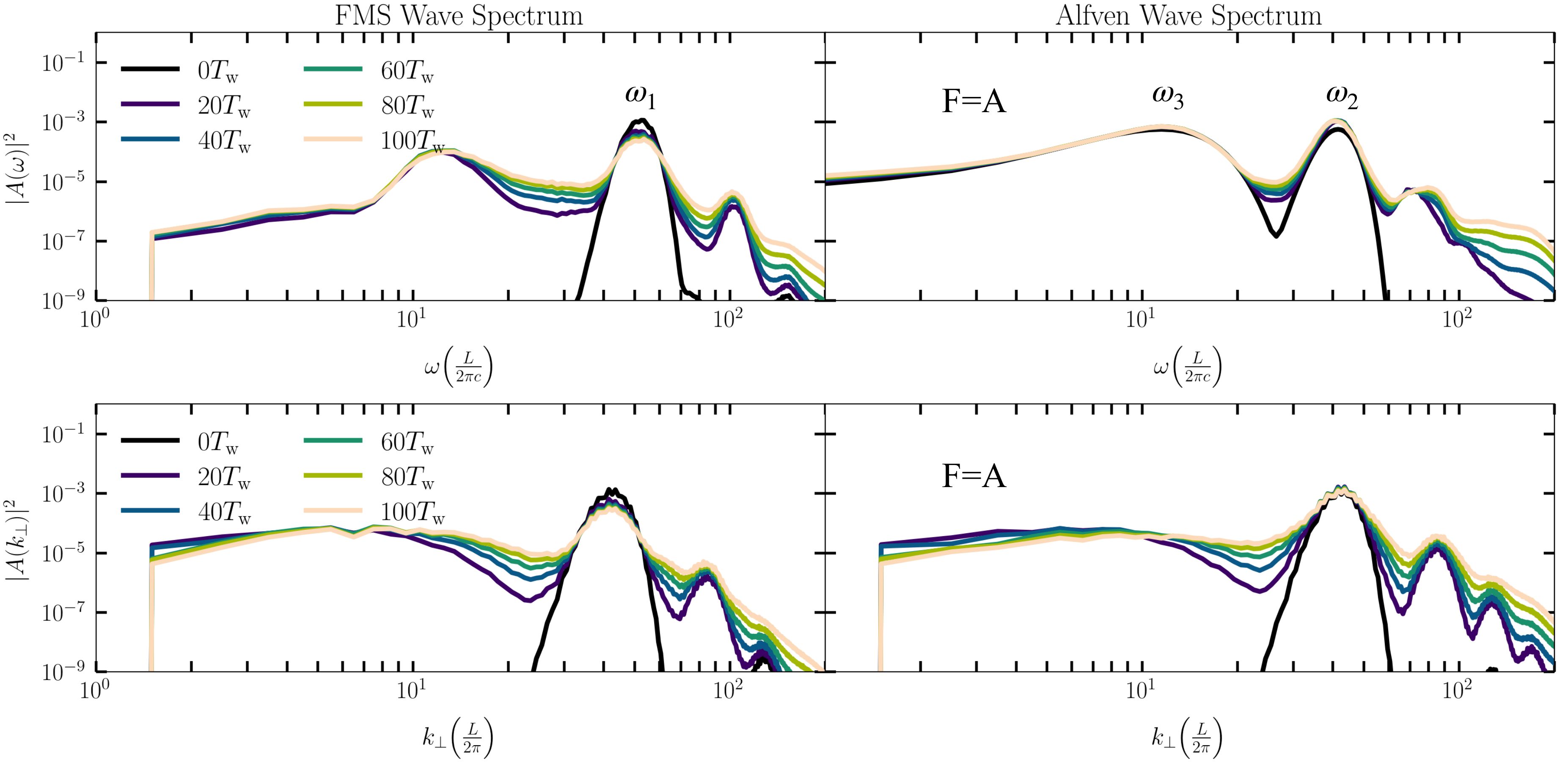}
    \includegraphics[width=0.98\textwidth]{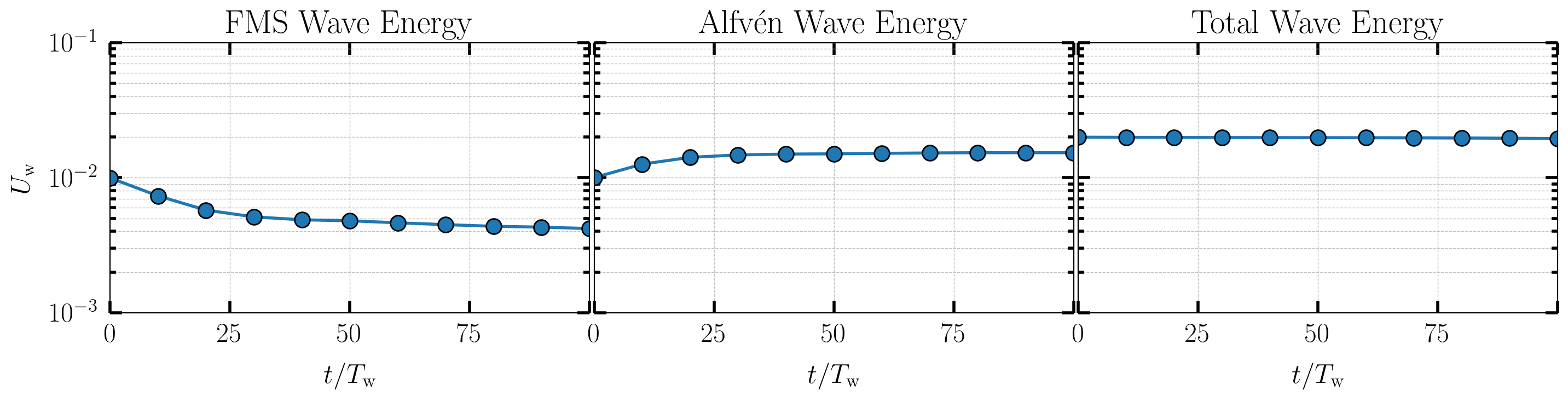}
    \caption{Time evolution of the simulation with a strong Alfvénic background (``\texttt{F=A}''), in which the initial Alfvén wave (AW) energy equals that of the FMS wave. The top panel shows the frequency spectrum, the middle panel shows the $k_{\perp}$ distribution, and the bottom panel is analogous to Figure~\ref{fig:fw_wave_energies}. The AW spectrum develops an extended $k_{\perp}$ tail, similar to the ``\texttt{Mono-90}'' and ``\texttt{Broad}'' simulations, while the FMS wave energy steadily decreases over time, even when the AWs carry comparable energy.} 
    \label{fig:fwaw_wave_energies}
\end{figure*}

Figure~\ref{fig:fwaw_wave_energies} shows the time evolution of FMS, Alfvén, and total wave energies. Notably, the total FMS wave energy decreases while the AW energy continues to increase. These results indicate that approximate energy equipartition between FMS and AWs is not a stable configuration. Instead, energy is continually transferred to the generation of Alfvén waves with an extended $k_{\perp}$-spectrum.

\section{Discussion and Conclusions} \label{section:Discussion}

In this paper, we performed direct numerical simulations of the parametric decay of a mildly non-linear spectrum of FMS waves via the $F \longleftrightarrow F+A$ and $F \longleftrightarrow A+A$ interactions. These interactions involve non-local energy transfer in $k$-space: AWs with large $k_{\perp}$ can be excited directly from FMS waves with smaller $k_{\perp}$. Our results are in agreement with theoretical predictions from \citet{lyubarsky_review_frb_2021Univ....7...56L} and spectral calculations presented in \citet{golbraikh_2023ApJ...957..102G}, which suggest that in the inner magnetar magnetosphere, FMS waves are efficiently converted into an Alfvénic continuum. We find that even for comparable energies of FMS and Alfvén waves, the parametric decay of the FMS wave remains the dominant interaction process. FRB-like signals produced deep in magnetar magnetospheres are therefore likely to suffer substantial energy loss before escaping.

Low-frequency electromagnetic waves can lose energy through several mechanisms as they propagate through magnetized plasma. In the magnetar context, these include: 
(i) an enhanced scattering cross-section for waves whose amplitudes exceed the background magnetic field \citep{beloborodov_khz_2021ApJ...922L...7B, Lyutikov2021:2110.08435v1, Qu2022:2204.10953v3, Beloborodov2023:2307.12182v2, Huang2024:2410.04065v1, Nishiura2024:2411.00936v1}, 
(ii) the formation of shocks when the electric field approaches the strength of the magnetic field \citep{Beloborodov_2023arXiv230712182B, jens_magnetar_2024ApJ...972..139M, bernardi_magnetar_2025ApJ...980..222B, Vanthieghem_levinson_ghz2024:2407.15076v1}, and 
(iii) parametric decay via weak turbulence \citep{lyubarsky_review_frb_2021Univ....7...56L, golbraikh_2023ApJ...957..102G}, which is the focus of this paper. Among these processes, weak turbulence is particularly notable because it can operate even when the FMS wave is only mildly non-linear, transferring energy into Alfvén waves via the $F \longleftrightarrow A+A$ process.\footnote{It is additionally insensitive to the plasma velocity along the magnetic field, as EM field dynamics in FFE are unaffected by parallel plasma motion, provided sufficient current is supplied.} This poses a significant challenge for near-field FRB models, which rely on the propagation of FMS waves through the inner magnetosphere.

Our simulations were performed in a periodic box to allow direct comparison with existing analytical work. In realistic magnetospheres, significant amplification of low-amplitude background AWs can occur only if they spatially overlap with the outgoing FMS pulse. Since the group velocity of Alfvén waves is aligned with the background magnetic field and approaches the speed of light, the interaction time is limited. In a companion paper (Solanki et al., in preparation), we confirm that the conclusions drawn in this work remain valid for the propagation of more realistic FMS pulses with finite envelope lengths.

\section{Acknowledgments}
We thank Ben Chandran, Amir Levinson, Yuri Lyubarsky, Anatoly Spitkovsky, and Chris Thompson for fruitful discussions. SS acknowledges helpful comments from the audience at the CCA SPA and the Princeton Plasma Astrophysics meetings. This work was supported by NSF grant No. AST-2307395 (AP), the Simons Foundation (00001470, SS and AP), and facilitated by the Multimessenger Plasma Physics Center (MPPC, AP), NSF grant No. PHY-2206607. A.P. additionally acknowledges support from an Alfred P. Sloan Fellowship and a Packard Foundation Fellowship in Science and Engineering. J.F.M. acknowledges support from NSF grant AST-2508744. This research is part of the Frontera computing project \citep{Frontera} at the Texas Advanced Computing Center (LRAC-AST21006), which is made possible by NSF award OAC-1818253.

\appendix

\section{Three Wave Interaction Matrix Coefficients } \label{appendix:three_wave_interactions}
We derive the interaction coefficients for the $F \longleftrightarrow F+A$ and $F \longleftrightarrow A+A$ mode coupling in FFE (similar derivations have been done in the Appendices of \cite{Lyubarsky_crab_2019MNRAS.483.1731L,Li_beloborodov_2019ApJ...881...13L}). As the simplest case, we consider a system with an FMS wave with wavevector $\bfm{k}_1$, an Alfvén wave with a wavevector $\bfm{k}_2$, and a third mode, which can either be an FMS or an Alfvén wave, with a wavevector $\bfm{k}_3$. We will further assume the three waves satisfy the resonance conditions $\bfm{k}_1 = \bfm{k}_2 + \bfm{k}_3$ and $\omega_{\bfm{k}_1} = \omega_{\bfm{k}_2} + \omega_{\bfm{k}_3}$. 
\newline
Combining Ampere's and Faraday's Laws, we can write a wave equation for the total electric field:
\begin{align}
    c^2 \curl{(\curl{\bfm{E}})} + \frac{\partial^2 \bfm{E}}{\partial t^2} = -4\pi \frac{\partial \bfm{J}}{\partial t},
    \label{eq:maxwell_wave}
\end{align}
which is supplemented by the force-free current:
\begin{align}
    \bfm{J} = \frac{c}{4 \pi B^2} \left[ \left(\dvg{E}\right) \Hquad \bfm{E} \times \bfm{B} + \bfm{B} \left(\bfm{B} \cdot (\curl{B}) - \bfm{E} \cdot (\curl{E}) \right) \right].
\end{align}
\newline
We perform the calculations in a quasi-linear fashion, where we expand the electric and magnetic fields up to second order. The zeroth-order background field is taken along the $z$-axis and has a magnitude $B_0$. The total electric and magnetic fields in the system are given by $\bfm{E} = \bfm{E}^{(1)} +  \bfm{E}^{(2)}$ and $\bfm{B} = B_0 \bfm{\hat{z}} + \bfm{B}^{(1)} + \bfm{B}^{(2)}$, respectively. 
The fields of the three plane waves sum up to $\bfm{E}^{(1)}$  and $\bfm{B}^{(1)}$, where each wave has a slowly amplitude and phase given by phasors $\chi_{\bfm{k}'}(t)$ and polarizations $\bfm{\delta \hat{E}}(\bfm{k}') $ and $\bfm{\delta \hat{B}}(\bfm{k}')$:
\begin{align}
    \bfm{E}^{(1)} &= \sum_{\bfm{k}' = {\bfm{k}_1,\bfm{k}_2, \bfm{k}_3}} \chi_{\bfm{k'}}(t) e^{i(\bfm{k}' \cdot \bfm{r} - \omega_{\bfm{k}'}t)} \bfm{\delta \hat{E}}(\bfm{k}') + \mathrm{c.c.} \\
    \bfm{B}^{(1)} &= \sum_{\bfm{k}' = {\bfm{k}_1,\bfm{k}_2, \bfm{k}_3}} \chi_{\bfm{k'}}(t)  e^{i(\bfm{k}' \cdot \bfm{r} - \omega_{\bfm{k}'}t)} \bfm{\delta \hat{B}}(\bfm{k}') + \mathrm{c.c.}.
\end{align}
The polarizations for FMS and Alfvén waves with a wavevector $\bfm{k}'$ are given by:
\begin{align}
    \delta \bfm{\hat{E}}^{(1)}_{\rm F}(\bfm{k'}) &= \bfm{\hat{k}'}_{\perp} \times \bfm{\hat{z}} \\
    \delta  \bfm{\hat{B}}^{(1)}_{\rm F}(\bfm{k'}) &= \frac{k'_z}{k'}\bfm{\hat{k}'}_{\perp} - \frac{k'_{\perp}}{k'}\bfm{\hat{z}}  \\
    \delta \bfm{\hat{E}}^{(1)}_{\rm A}(\bfm{k'}) &= \bfm{\hat{k}'}_{\perp} \qquad \\
    \delta  \bfm{\hat{B}}^{(1)}_{\rm A}(\bfm{k'}) &= \mathrm{sign}(k'_z) \bfm{\hat{z}} \times \bfm{\hat{k}'}_{\perp} \qquad.
\end{align}
Additionally, the $\bfm{\hat{z}}$ component of $\bfm{E}^{(2)}$ is constrained from the $\bfm{E} \cdot \bfm{B} = 0$ condition. That is, 
\begin{align}
    \bfm{E}^{(2)} \cdot \bfm{\hat{z}} = -\frac{\bfm{E}^{(1)} \cdot \bfm{B}^{(1)}}{B_0}
    \label{eq:e_second_order}
\end{align}

Expanding the non-linear current using the electric and magnetic fields, and keeping only the second order terms, the evolution equation for the electric field in Eq. (\ref{eq:maxwell_wave}) now becomes:
\begin{align}
    c^2 (\nabla\times (\nabla\times {\bfm{E}^{(2)}})) + \frac{\partial^2 \bfm{E}^{(2)}}{\partial t^2} - \sum_{\bfm{k'}} \left( 2i\omega_{\bfm{k'}} \frac{\partial \chi_{\bfm{k'}}}{\partial t} e^{i(\bfm{k}' \cdot \bfm{r} - \omega_{\bfm{k}'}t)} \bfm{\delta \hat{E}(\bfm{k'})} + \mathrm{c.c.} \right)
    = -4 \pi \frac{\partial \bfm{J}^{(2)}}{\partial t},
    \label{eq:linearized_wave_eq}
\end{align}
where the sum runs over $\bfm{k}_1, \bfm{k}_2$ and $\bfm{k}_3$, and the second-order non-linear current is given by:
\begin{equation}
\begin{aligned}
\bfm{J}^{(2)} = \frac{c}{4\pi B_0} \Big[ 
& (\nabla \cdot \bfm{E}^{(1)}) (\bfm{E}^{(1)} \times \bfm{\hat{z}}) 
+ (\bfm{\hat{z}} \cdot (\nabla \times \bfm{B}^{(1)})) \bfm{B}^{(1)} \\[4pt]
& + \bfm{\hat{z}} \Big( 
    \bfm{B}^{(1)} \cdot (\nabla \times \bfm{B}^{(1)}) 
    - \bfm{E}^{(1)} \cdot (\nabla \times \bfm{E}^{(1)}) \\[4pt]
& \quad - 2 (\bfm{B}^{(1)} \cdot \bfm{\hat{z}}) (\nabla \times \bfm{B}^{(1)})_z 
    + B_0 (\nabla \times \bfm{B}^{(2)})_z 
\Big)
\Big].
\label{eq:second_order_nl_current}
\end{aligned}
\end{equation}
We want to solve for $\frac{ \partial \chi_{\bfm{k'}}}{\partial t}$ in Eq. (\ref{eq:linearized_wave_eq}) by projecting the equation along $\bfm{\delta \hat{E}}(\bfm{k}')$, and only keeping the terms that resonate with $e^{i(\bfm{k'\cdot r} - \omega_{\bfm{k'}}t)}$ for $\bfm{k}' = \bfm{k}_1, \bfm{k}_2$ and $\bfm{k}_3$. 

Consider first, the $  F\longleftrightarrow F+A$ process, where $\bfm{k}_3$ is an FMS mode. 
First, projecting Eq. (\ref{eq:linearized_wave_eq}) along $\bfm{\delta \hat{E}}(\bfm{k}_1)$, using the FMS dispersion relation $\omega_{\bfm{k}_1} = |\bfm{k}_1|c$, and picking out only the resonant terms, we get:
\begin{align}
    \frac{\partial \chi_{\bfm{k}_1}}{\partial t} = \frac{i c k_{2,\perp}}{2 B_0} \left[\bfm{\hat{z}} \cdot (\bfm{\hat{k}_{3,\perp}} \times \bfm{\hat{k}_{1,\perp}}) \cdot \left(\frac{k_{3,z}}{k_3} \mathrm{sign}(k_{2,z} ) - 1 \right) \right] \chi_{\bfm{k}_2} \bfm{\chi_{k_3}}.
    \label{eq:ffa_k}
\end{align}
Similarly, projecting along $\bfm{\delta \hat{E}}(\bfm{k_2})$ and making use of Eq. (\ref{eq:e_second_order}), we get an equation for Alfvén wave growth:
\begin{align}
    \frac{\partial \chi_{\bfm{k}_2}}{\partial t} = \frac{-i c k_{2,\perp} \mathrm{sign}(k_{2,z})}{2 B_0} \left[\bfm{\hat{z}} \cdot (\bfm{\hat{k}_{3,\perp}} \times \bfm{\hat{k}_{1,\perp}}) \cdot \left(\frac{k_{3,z}}{k_3}  - \frac{k_{1,z}}{k_1} \right) \right] \chi_{\bfm{k}_1} \bfm{\chi_{k_3}^*}.
    \label{eq:ffa_k1}
\end{align}
Finally, projecting along $\bfm{\delta \hat{E}}(\bfm{k_3})$, we obtain:
\begin{align}
    \frac{\partial \chi_{\bfm{k}_3}}{\partial t} = \frac{-i c k_{2,\perp}}{2 B_0} \left[\bfm{\hat{z}} \cdot (\bfm{\hat{k}}_{1,\perp} \times \bfm{\hat{k}}_{3,\perp}) \cdot \left(\frac{k_{1,z}}{k_1} \mathrm{sign}(k_{2,z} ) - 1 \right) \right] \chi_{\bfm{k}_1} \bfm{\chi_{k_2}^*}.
    \label{eq:ffa_k2}
\end{align}
Making the transformation $a_{\bfm{k'}} = \chi_{\bfm{k'}} e^{i\omega_{\bfm{k'}t}} / \sqrt{2 \pi \omega_{\bfm{k'}}}$ for $\bfm{k}' = \bfm{k}_1, \bfm{k}_2$ and $\bfm{k}_3$, we can express the equations as:
\begin{align}
    \left(\frac{\dd }{\dd t} + i\omega_{\bfm{k}_1}\right) a_{\bfm{k}_1}   &= iV_{\rm FFA} a_{\bfm{k}_2} a_{\bfm{k}_3} \\
    \left(\frac{\dd }{\dd t} + i\omega_{\bfm{k}_2}\right) a_{\bfm{k}_2} &= iV_{\rm FFA} a_{\bfm{k}_1} a_{\bfm{k}_3}^* \\
    \left(\frac{\dd }{\dd t} + i\omega_{\bfm{k}_3} \right) a_{\bfm{k}_3}   &= iV_{\rm FFA} a_{\bfm{k}_1} a_{\bfm{k}_2}^*,
\end{align}
which are of the same form as Eqs. (\ref{eq:three_wave_rescaled_1})-(\ref{eq:three_wave_rescaled_3}), with the $V_{\rm FFA}$ given by Eq. (\ref{eq:vffa}). Similarly, the matrix coefficient for the $V_{\rm FAA}$ can be calculated by projecting Eq. (\ref{eq:linearized_wave_eq}) along the electric fields for the FMS and the two AWs. For example, the evolution equation for the FMS phasor $\chi_{\bfm{k}_1}$ becomes:

\begin{equation}
    \frac{\partial \chi_{\bfm{k}_1}}{\partial t} = \frac{ic k_{2, \perp}}{2 B_0} (1 - \mathrm{sign}(k_{2,z} k_{3,z})) \Bigl[(\bfm{\hat{k}_{1,\perp}} \cdot \bfm{\hat{k}_{3,\perp}}) - \frac{k_{3,\perp}}{k_{2,\perp}} (\bfm{\hat{k}_{1,\perp}} \cdot \bfm{\hat{k}_{2,\perp}}) \Bigr] \chi_{\bfm{k}_2} {\chi_{\bfm{k}_3}}
\end{equation}

Similarly, the equations for  $\chi_{\bfm{k}_2}$ and  $\chi_{\bfm{k}_3}$ can be obtained. The corresponding matrix coefficient is given in Eq. (\ref{eq:vfaa}). 

\section{Momentum Conservation} \label{appendix:momentum conservation}
The total momentum in the product FMS and AWs in three-wave interactions is insufficient to account for the total momentum of the initial FMS wave. However, FFE maintains momentum conservation by creating a second-order, non-linear electric field $\bfm{\bar{E}^{(2)}}$, so that $\bfm{S}_{\rm NL} \equiv (c/4\pi) \, \bfm{\bar{E}^{(2)}} \times \bfm{B_0}$ accounts for the change in perpendicular momentum of the decaying FMS wave. This electric field has a spatially uniform component $\bfm{k} = 0$, and is produced via non-linear currents at the third order\footnote{Third-order currents are necessary to get a zero-wavevector term via the wavevector matching condition. That is, the terms of the form $A_1  A_2 A_3 e^{i(\bfm{(k_1-k_2-k_3) \cdot r})}$ in the non-linear current, where $A_i \equiv |\chi_{\bfm{k}_i}|$ are the wave amplitudes, will produce $\bfm{k}=0$ electric fields.}.

\begin{figure}
    \centering
    \includegraphics[width=0.5\textwidth]{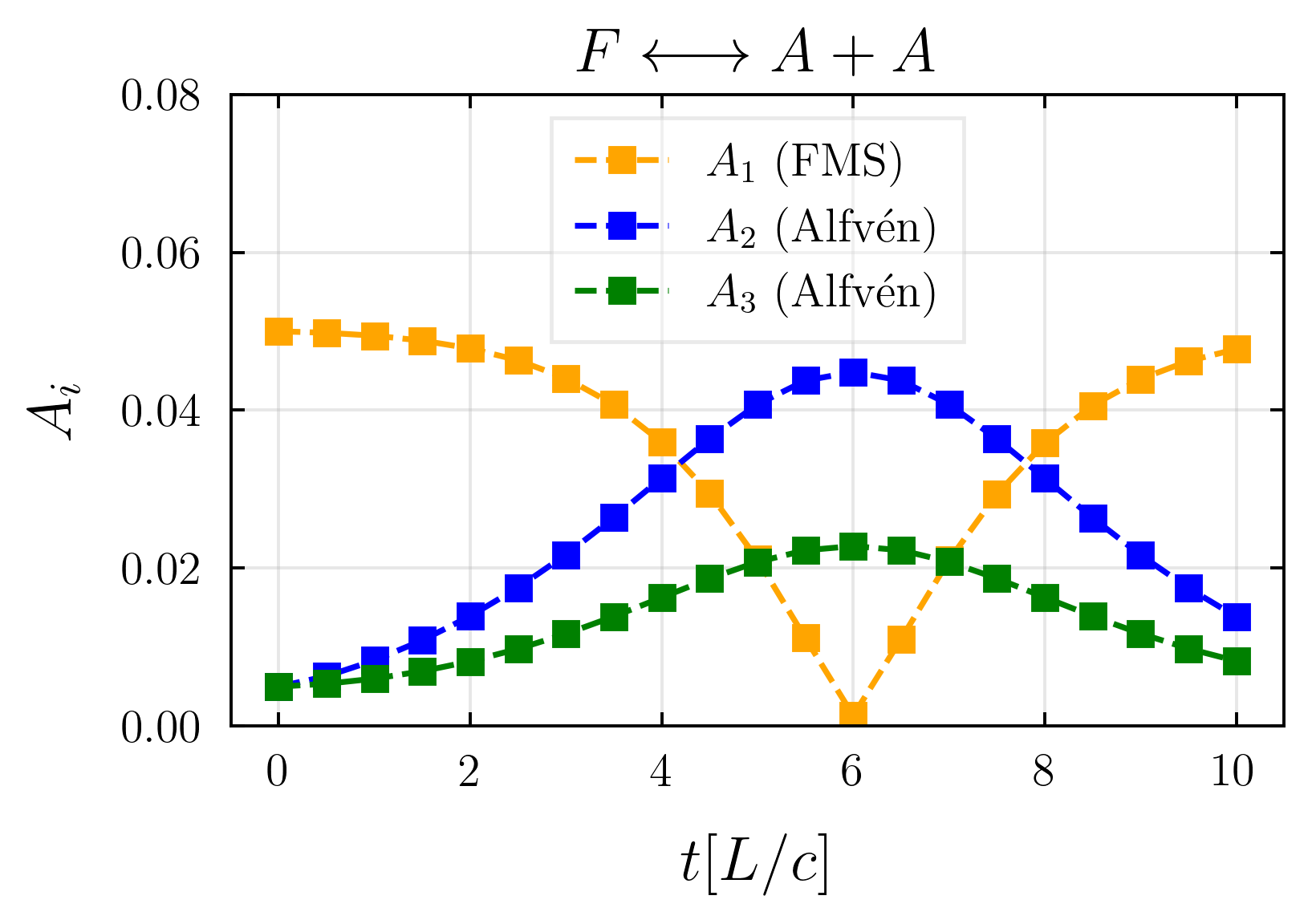}
    \includegraphics[width=0.99\textwidth]{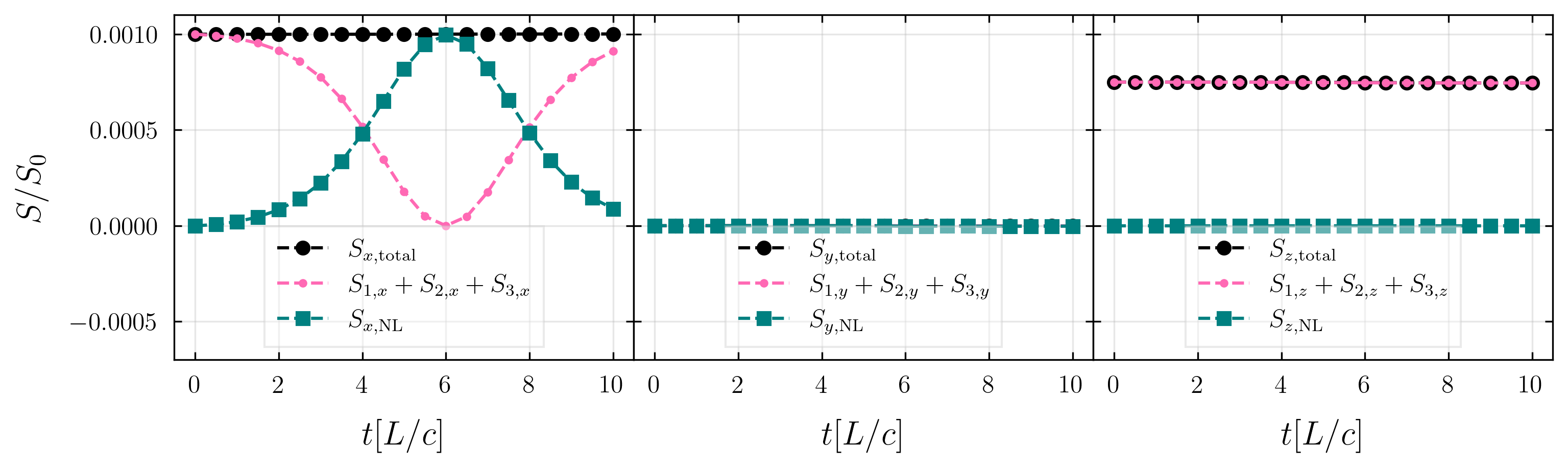}
    \caption{(Top): The amplitudes of the FMS and two AWs in the $F \longleftrightarrow A+A$ simulation. The two AWs are initialized with a much smaller amplitude than the FMS wave (but a higher amplitude than the background noise so these modes grow the fastest). (Bottom): Normalized Poynting flux decomposition from the same simulation, where $S_0 = (c/4\pi)B_0^2$. The three panels represent the volume-averaged Poynting flux in the $x, y$, and $z$ directions, respectively. The black curve represents the total Poynting flux, $\bfm{S}_{\rm tot}$, in the simulation, which is conserved along each of the directions. The pink curve represents the sum of the Poynting fluxes for the FMS and two AWs, given by $\bfm{S}_1 + \bfm{S}_2 + \bfm{S}_3$. The perpendicular ($x$) component of this flux is not conserved; when the FMS decays, this flux also goes to zero. The teal curve represents the non-linear Poynting flux, $\bfm{S}_{\rm NL}$, which is generated by non-linear electric fields. $\bfm{S}_{\rm NL}$ accounts for the perpendicular component of $\bfm{S}_1$ when the FMS wave decays, so that $\bfm{S}_{\rm NL} + \bfm{S}_1 + \bfm{S}_2 + \bfm{S}_3 = \bfm{S}_{\rm tot}$ is a constant. }
    \label{fig:poynting_flux}
\end{figure}

We demonstrate this by running an FFE simulation of the $F \longleftrightarrow A+A$ process. We initialize the simulation with an FMS and two AWs with parameters in Table \ref{tab:waves_amplitudes}. These waves satisfy the resonance conditions (\ref{eq:resonance_k}) and (\ref{eq:resonance_omega}). The top panel in Fig. \ref{fig:poynting_flux} shows the evolution of the amplitudes of the three waves over time. The FMS wave decays after time $t \sim 6 \, L/c$ and grows the two AWs at the expense of its energy.

\begin{table}[t]
\centering
\begin{tabular}{lcc}
\hline
Wavetype & Wavevector $(k_x,k_y,k_z)$ & Amplitude \\
\hline
FMS    & $\frac{2\pi}{L}(8,0,6)$  & 0.05 \\
Alfvén & $\frac{2\pi}{L}(6,-3,8)$ & 0.005 \\
Alfvén & $\frac{2\pi}{L}(2,3,-2)$ & 0.005 \\
\hline
\end{tabular}
\caption{Parameters of the three waves in the FFE simulation used to verify momentum conservation.}
\label{tab:waves_amplitudes}
\end{table}

Each column in the second row of Fig. \ref{fig:poynting_flux} represents the $x, y$ and $z$ components of the Poynting flux. We plot the total Poynting flux, $\bfm{S}_{\rm tot}$ (black curve), the Poynting flux from the sum of the three waves, $\bfm{S}_1 + \bfm{S}_2 + \bfm{S}_3$ (pink curve) and the non-linear Poynting flux, $\bfm{S}_{\rm NL}$  (teal curve), which is computed as $(c / 4\pi) \bfm{\bar{E}^{(2)}} \times \bfm{B}_0$, where $\bfm{\bar{E}^{(2)}}$ is the volume averaged electric field. The growth in $\bfm{S}_{\rm NL, \perp}$ balances the decrease in $\bfm{S}_{1,\perp}$ as the FMS wave decays. A non-linear Poynting flux is required because $\bfm{S}_{2,3}$ lie along the background magnetic field, and cannot account for the flux in the perpendicular direction. $\bfm{S}_{\rm NL}$ is produced from a non-linear electric field with a $\bfm{k}=(0,0,0)$ component. In the remainder of this section, we show how such an electric field is generated in a system that starts out with just the three linear waves.

There are a few steps involved in the generation of $\bfm{\bar{E}^{(2)}}$: initially, first-order linear waves interact and produce non-linear second-order electric and magnetic fields $\bfm{E^{(2)}}$ and $\bfm{B^{(2)}}$ with $\omega \neq 0$. These second-order fields, in conjunction with the linear waves, produce a third-order FFE current, $\bfm{J^{(3)}}$, which has a $\omega = 0$ and $\bfm{k} = (0,0,0)$ component, $\bfm{\bar{J}^{(3)}}$. This current produces an electric field with a corresponding $\omega=0$ and $\bfm{k}=(0,0,0)$ component that grows in amplitude over the long-timescale $t^{(1)}_{\rm int}$ and becomes second-order in wave-amplitudes: $\bfm{\bar{E}^{(2)}} = - (4\pi / c)\int_0 ^{t^{(1)}_{\rm int}} \bfm{\bar{J}^{(3)}} dt$. It is the Poynting flux from $\bfm{\bar{E}^{(2)}}$ that balances out the missing Poynting flux from the initial FMS wave. 

We will first compute $\bfm{E^{(2)}}$ and $\bfm{B^{(2)}}$. Using Eqs. (\ref{eq:maxwell_wave}), (\ref{eq:e_second_order}) and (\ref{eq:second_order_nl_current}), we get the following equation for the evolution of the second-order electric field:
\begin{equation}
    c^2 \left[\curl{\curl{E^{(2)}}}\right]_{\perp} + \frac{\partial^2 \bfm{E^{(2)}_{\perp}}}{\partial t^2} = -4\pi \frac{\partial \bfm{J^{(2)}_{\perp}}}{\partial t}.
    \label{eq:second_order_wave_eqn}
\end{equation}
In the above equation, $\bfm{J^{(2)}_{\perp}}$ is the perpendicular component of the non-linear current that is entirely given by the first-order electric and magnetic fields. It takes the form
\begin{equation}
    \bfm{J^{(2)}_{\perp}} = \frac{c}{4 \pi B_0} \left[\rho^{(1)} \bfm{E^{(1)}} \times \bfm{\hat{z}} + \bfm{B^{(1)}_{\perp}} J^{(1)}_z  \right].
    \label{eq:second_order_current}
\end{equation}
Using the FMS and AW polarizations, we can write the electric and magnetic fields for first-order linear waves: $\bfm{E^{(1)}} = \bfm{E}_{\rm F1} + \bfm{E}_{\rm A2} + \bfm{E}_{\rm A3}$ and $\bfm{B^{(1)}} = \bfm{B}_{\rm F1} + \bfm{B}_{\rm A2} + \bfm{B}_{\rm A3}$, where:
\begin{align}
    \bfm{E}_{\rm F1} &= A_1 \cos(\bfm{k_1 \cdot r} - \omega_1 t + \phi_1) (\bfm{\hat{k}_{1, \perp} \times \hat{z}}) \label{eq:e_f1}\\
    \bfm{E}_{\rm A2} &= A_2 \cos(\bfm{k_2 \cdot r} - \omega_2 t + \phi_2) (\bfm{\hat{k}_{2, \perp}}) \label{eq:e_a2}\\
    \bfm{E}_{\rm A3} &= A_3 \cos(\bfm{k_3 \cdot r} - \omega_3 t + \phi_3) (\bfm{\hat{k}_{3, \perp}}) \label{eq:e_a3} \\
    \bfm{B}_{\rm F1} &= A_1 \cos(\bfm{k_1 \cdot r} - \omega_1 t + \phi_1) \Bigl(\frac{k_{1,z}}{k_1} \bfm{\hat{k}_{1,\perp}} - \frac{k_{1,\perp}}{k_1} \bfm{\hat{z}} \Bigr) \label{eq:b_f1}\\
    \bfm{B}_{\rm A2} &= A_2 \cos(\bfm{k_2 \cdot r} - \omega_2 t + \phi_2) \mathrm{sign}(k_{2,z}) (\bfm{\hat{z}} \times \bfm{\hat{k}_{2, \perp}}) \label{eq:b_a2} \\
    \bfm{B}_{\rm A3} &= A_3 \cos(\bfm{k_3 \cdot r} - \omega_3 t + \phi_3) \mathrm{sign}(k_{3,z}) (\bfm{\hat{z}} \times \bfm{\hat{k}_{3, \perp}}) \label{eq:b_a3}
\end{align}
Here $A_i \equiv |\chi_{\bfm{k}_i}|$ is the absolute value of the amplitude, following definition of $\chi_{\bfm{k}_i}$ in Appendix \ref{appendix:three_wave_interactions}. 
Without the loss of generality, we can take $\bfm{k_1} = (k_{1,x}, 0, k_{1,z})$, $k_{2,z} >0$ and $k_{3,z} < 0$ (this is the same setup we have in the $F \longleftrightarrow A+A$ simulation -- see Table \ref{tab:waves_amplitudes} for simulation parameters). The $z-$component of $\bfm{E^{(2)}}$ is fixed from the $\bfm{E} \cdot \bfm{B} = 0$ condition in Eq. (\ref{eq:e_second_order}). Namely,
\begin{align}
E^{(2)}_z = -\frac{\mathbf{E}^{(1)} \cdot \mathbf{B}^{(1)}}{B_0}
= -\frac{1}{4 B_0} \left[ 
    A_1A_2 \alpha_{1,2}\!\left(e^{i\xi_{1,2}^+} + e^{i\xi_{1,2}^-}\right)
  + A_1A_3 \alpha_{1,3}\!\left(e^{i\xi_{1,3}^+} + e^{i\xi_{1,3}^-}\right) 
  + 2 A_2A_3 \alpha_{2,3}\!\left(e^{i\xi_{2,3}^+} + e^{i\xi_{2,3}^-}\right)
  + \text{c.c.}
  \label{eq:e2_z}
\right],
\end{align}
where $\xi_{i,j}^{\pm} = (\bfm{k_i \pm k_j}) \cdot \bfm{r} - (\omega_i \pm \omega_j)t + \phi_{i} \pm \phi_j$, $\alpha_{1,2} = \cos \Phi_2 (\cos \theta_1 - 1)$, $\alpha_{1,3} = \cos \Phi_3 (1 + \cos \theta_1)$ and $\alpha_{2,3} = (\cos \Phi_2 \sin \Phi_3 - \sin \Phi_2 \cos \Phi_3)$, with $\cos \Phi_{i} = k_{i,x} / k_{i, \perp}, \sin \Phi_{i} = k_{i,y} / k_{i, \perp}$ and $\cos \theta_i = k_{i,z} / k_i$ .  

The perpendicular component of the second-order electric field obtained using Eq. (\ref{eq:second_order_wave_eqn}), along with the  $z$-component from Eq. (\ref{eq:e2_z}), gives the total electric field in the system. We will now solve for this second-order electric field. 

Taking wavelike solutions for the electric field of the form $\bfm{E^{(2)}_{\perp}} = (2 \pi)^{-3} \int (\bfm{\hat{E}^{(2)}}(\bfm{k}, t) e^{i \bfm{k\cdot r}} + \mathrm{c.c.}) d\bfm{k}$ and for $E^{(2)}_z$ and $\bfm{J^{(2)}_{\perp}}$,  Eq. (\ref{eq:second_order_wave_eqn}) can be projected along $\bfm{\hat{E}^{(2)}}(\bfm{k})$ along $\bfm{\hat{k}_{\perp} \times \hat{z}}$ and $\bfm{\hat{k}_{\perp}}$, respectively:

\begin{align}
    \Bigl(k^2c^2 + \frac{\partial^2}{\partial t^2} \Bigr) \hat{E}^{(2)}_{\hat{k}_{\perp} \times \hat{z}} &= -4\pi \frac{\partial }{\partial t}  \hat{J}^{(2)}_{\hat{k}_{\perp} \times \hat{z}} \label{second_order_fms_pol} \\
    \Bigl(k_z^2c^2 + \frac{\partial^2}{\partial t^2} \Bigr) \hat{E}^{(2)}_{\hat{k}_{\perp}} &= -4\pi \frac{\partial }{\partial t}  \hat{J}^{(2)}_{\hat{k}_{\perp} } + k_{\perp}k_z c^2 \hat{E}^{(2)}_z  \label{second_order_aw_pol} .
\end{align}
 We take the initial conditions to be $\bfm{\hat{E}_{(2)}}(\bfm{k},t=0) = 0$ and $\frac{\partial }{\partial t}\bfm{\hat{E}_{(2)}}(\bfm{k},t=0) = 0$. Note that Eqs. (\ref{second_order_fms_pol})-(\ref{second_order_aw_pol}) are simply equations for driven harmonic oscillators - for the $\bfm{k}$'s where the RHS is zero, no electric field gets excited, given the initial conditions. 

To proceed, we need to evaluate the time derivative of the second-order current. Substituting Eqs. (\ref{eq:e_f1})-(\ref{eq:b_a3}) in Eq. (\ref{eq:second_order_current}), the second-order current becomes:

\begin{equation}
    \bfm{J^{(2)}_{\perp}} = \frac{-ic}{16 \pi B_0} \left[ A_1 A_2 \,\vec{\boldsymbol{\beta}}_{1,2} (e^{i\xi_{1,2}^+} - e^{i\xi_{1,2}^-}) + A_1 A_3 \,\vec{\boldsymbol{\beta}}_{1,3} (e^{i\xi_{1,3}^+} - e^{i\xi_{1,3}^-}) + A_2 A_3 (\,\vec{\boldsymbol{\beta}}_{2,3} (e^{i\xi_{2,3}^+} - e^{i\xi_{2,3}^-})
    + \,\vec{\boldsymbol{\beta}}_{3,2} (e^{i\xi_{3,2}^+} - e^{i\xi_{3,2}^-})) 
    \right] +  \mathrm{c.c.},
    \label{j_second_order}
\end{equation}

where $\vec{\boldsymbol{\beta}}_{1,2} = k_{2,\perp}(1 - \cos \theta_1) \bfm{\hat{k}_{1, \perp}}, \vec{\boldsymbol{\beta}}_{1,3} = k_{3,\perp}(1 + \cos \theta_1) \bfm{\hat{k}_{1, \perp}}, \vec{\boldsymbol{\beta}}_{2,3} = -2 k_{3, \perp} (\bfm{\hat{k}_{2, \perp} \times \hat{z}})$ and $ \vec{\boldsymbol{\beta}}_{3,2} = -2 k_{2, \perp} (\bfm{\hat{k}_{3, \perp} \times \hat{z}})$. Its time derivative is
\begin{equation}
\begin{split}
\frac{\partial}{\partial t} \mathbf{J}^{(2)}_{\perp}
  &= -\frac{c}{16 \pi B_0} \Biggl[
     A_1 A_2 \,\vec{\boldsymbol{\beta}}_{1,2}
       \left(e^{i\xi_{1,2}^+} \omega^{+}_{1,2}
           - e^{i\xi_{1,2}^-}\omega^{-}_{1,2}\right)  
 + A_1 A_3 \,\vec{\boldsymbol{\beta}}_{1,3}
       \left(e^{i\xi_{1,3}^+} \omega^{+}_{1,3}
           - e^{i\xi_{1,3}^-}\omega^{-}_{1,3}\right)  \\
  &\quad + A_2 A_3 \Biggl\{
        \vec{\boldsymbol{\beta}}_{2,3}
          \left(e^{i\xi_{2,3}^+}\omega^{+}_{2,3}
              - e^{i\xi_{2,3}^-}\omega^{-}_{2,3}\right)
      + \vec{\boldsymbol{\beta}}_{3,2}
          \left(e^{i\xi_{3,2}^+}\omega^{+}_{3,2}
              - e^{i\xi_{3,2}^-}\omega^{-}_{3,2}\right)
     \Biggr\}
   \Biggr] + \mathrm{c.c.},
\end{split}
\label{j_second_order_2}
\end{equation}
with $\omega^{\pm}_{i,j} \equiv \omega_{i} \pm \omega_j$.

To evaluate Eqs. (\ref{second_order_fms_pol}) - (\ref{second_order_aw_pol}), we need to express $E^{(2)}_z$ and $\tfrac{\partial}{\partial t} \bfm{J^{(2)}_{\perp}}$ in Fourier space. Note that both of these contain terms that go as $e^{i \xi^{\pm}}$, where all the spatial dependence is stored. Defining $\bfm{k}^{\pm}_{i,j} \equiv \bfm{k}_i \pm \bfm{k}_j$ and $\eta^{\pm}_{i,j} \equiv \xi^{\pm}_{i,j} - \bfm{k}^{\pm}_{i,j} \cdot \bfm{r} = \omega^{\pm}_{i,j}t + (\phi_i \pm \phi_j)$ and using the relation $\int e^{i \xi^{\pm}_{i,j}} e^{-i \mathbf{k} \cdot \bfm{r}} d\bfm{r} = (2 \pi)^3 e^{i\eta^{\pm}_{i,j}} \delta (\bfm{k} - \bfm{k}^{\pm}_{i,j})$, Equations (\ref{second_order_fms_pol}) - (\ref{second_order_aw_pol}) become:

\begin{align}
    \bigl(k^2c^2 + \tfrac{\partial^2}{\partial t^2}\bigr) 
    \hat{E}^{(2)}_{\hat{k}_{\perp} \times \hat{z}} 
    &= \frac{(2\pi)^3 c}{4 B_0} \Biggl[
         A_1 A_2 \,\vec{\boldsymbol{\beta}}_{1,2}
       \Biggl\{\delta(\mathbf{k - k^+_{1,2}})\, e^{i\eta_{1,2}^+}\,\omega^{+}_{1,2}
           - \delta(\mathbf{k - k^-_{1,2}})\, e^{i\eta_{1,2}^-}\,\omega^{-}_{1,2}\Biggr\}   \notag \\
 &\quad+ A_1 A_3 \,\vec{\boldsymbol{\beta}}_{1,3}
       \Biggl\{\delta(\mathbf{k - k^+_{1,3}})\, e^{i\eta_{1,3}^+}\,\omega^{+}_{1,3}
           - \delta(\mathbf{k - k^-_{1,3}})\, e^{i\eta_{1,3}^-}\,\omega^{-}_{1,3}\Biggr\}   \notag \\
  &\quad + A_2 A_3 \Biggl\{
        \vec{\boldsymbol{\beta}}_{2,3}
          \left(\delta(\mathbf{k - k^+_{2,3}})\, e^{i\eta_{2,3}^+}\,\omega^{+}_{2,3}
              - \delta(\mathbf{k - k^-_{2,3}})\, e^{i\eta_{2,3}^-}\,\omega^{-}_{2,3}\right)  \notag \\
            &\quad
      + \vec{\boldsymbol{\beta}}_{3,2}
          \left(\delta(\mathbf{k - k^+_{3,2}})\, e^{i\eta_{3,2}^+}\,\omega^{+}_{3,2} 
              - \delta(\mathbf{k - k^-_{3,2}})\, e^{i\eta_{3,2}^-}\,\omega^{-}_{3,2}\right)
     \Biggr\} 
    \Biggr] \cdot (\mathbf{\hat{k}_{\perp} \times \hat{z}}) + \mathrm{c.c.} \label{eq:fms_fourier} \\
    (k_z^2c^2 + \tfrac{\partial^2}{\partial t^2}) \hat{E}^{(2)}_{\hat{k}_{\perp}} &= \frac{(2\pi)^3 }{4 B_0} \Biggl[
         A_1 A_2 \Biggl\{\delta(\bfm{k - k^+_{1,2}})  e^{i\eta_{1,2}^+}(-k_z k_{\perp} c^2 \alpha_{1,2} + c \vec{\boldsymbol{\beta}}_{1,2} \bfm{ \cdot \hat{k}_{\perp}} \omega^+_{1,2})
          \notag  \\ 
          &\quad - \delta(\bfm{k - k^-_{1,2}})  e^{i\eta_{1,2}^-}(k_z k_{\perp} c^2 \alpha_{1,2} + c \vec{\boldsymbol{\beta}}_{1,2} \bfm{ \cdot \hat{k}_{\perp}}\omega^-_{1,2})
         \Biggr\}
          \notag \\
         &\quad + A_1 A_3 \Biggl\{\delta(\bfm{k - k^+_{1,3}})  e^{i\eta_{1,3}^+}(-k_z k_{\perp} c^2 \alpha_{1,3} + c \vec{\boldsymbol{\beta}}_{1,3} \bfm{ \cdot \hat{k}_{\perp}}{\omega^+_{1,3}})
          \notag \\ 
          &\quad - \delta(\bfm{k - k^-_{1,3}})  e^{i\eta_{1,3}^-}(k_z k_{\perp} c^2 \alpha_{1,3} + c\vec{\boldsymbol{\beta}}_{1,3}\bfm{ \cdot \hat{k}_{\perp}} \omega^-_{1,3})
         \Biggr\}   \notag \\
         &\quad + A_2 A_3 \Biggl\{\delta(\bfm{k - k^+_{2,3}})  e^{i\eta_{2,3}^+} (-2 k_z k_{\perp} c^2 \alpha_{2,3} + c \omega^+_{2,3} (\vec{\boldsymbol{\beta}}_{2,3} + \vec{\boldsymbol{\beta}}_{3,2}) \cdot \bfm{\hat{k}_{\perp}}  )) 
           \notag \\ 
          &\quad - \delta(\bfm{k - k^-_{2,3}})  e^{i\eta_{2,3}^-} (2 k_z k_{\perp} c^2 \alpha_{2,3} + c \omega^-_{2,3} \vec{\boldsymbol{\beta}}_{2,3}   \cdot \bfm{\hat{k}_{\perp}}) -  \delta(\bfm{k - k^-_{3,2}})  e^{i\eta_{3,2}^-} c \omega^-_{3,2}
          \vec{\boldsymbol{\beta}}_{3,2}  \cdot \bfm{\hat{k}_{\perp}} \Biggr\} 
    \Biggr] + \mathrm{c.c.} \,.     \label{eq:aw_fourier}
\end{align}

The solutions are composed of resonant and non-resonant terms. The resonant solutions are simply the three-wave interactions -- these grow linearly with time (valid for $t \ll t_{\rm nl} \sim \omega_i^{-1} B_0 / B_i^{(1)}$). The non-resonant solutions oscillate at the natural and driving frequencies, given by the LHS frequencies ($k^2 c^2$ and $k_z^2 c^2$) and RHS frequencies ($\omega^{\pm}_{i,j}$ containted in $\eta^{\pm}_{i,j}$) of Eqs. (\ref{eq:fms_fourier})-(\ref{eq:aw_fourier}), respectively. We are interested in the non-resonant solutions with wavevectors $\bfm{k}_{\alpha}$ and frequencies $\omega_{\alpha}$, where $\alpha = \rm{1, 2, 3}$. These can interact with one of the three linear waves to produce a zero-frequency current. For example, the term containing $\delta (\bfm{k} - \bfm{k}^-_{1,2})$ produces an electric field with a wavevector $\bfm{k}^-_{1,2} = \bfm{k}_3$, which can interact with the linear AW with wavevector $\bfm{k}_3$ to produce a $\bfm{k}=0$ mode.  We will exclude all other terms that contain $\delta (\bfm{k} - \bfm{k}_{1,2}^+), \delta (\bfm{k} - \bfm{k}_{1,3}^+)$ or $\delta (\bfm{k} - \bfm{k}_{2,3}^-)$ because these cannot interact with a linear wave to produce a zero-frequency term. The terms of interest in the solutions of the above equations are:

\begin{align}
\hat{E}^{(2), \rm nres}_{\hat{k}_{\perp} \times \hat{z}} 
  &= -\frac{(2\pi)^3 c}{4 B_0} \Biggl[
     \frac{A_1 A_2 \,\delta(\mathbf{k - k^-_{1,2}})\, e^{i\eta_{1,2}^-}\,\omega^{-}_{1,2}}
          {c^2|\mathbf{k}_1 - \mathbf{k}_2|^2 - (\omega^-_{1,2})^2}
     \,\vec{\boldsymbol{\beta}}_{1,2} \cdot (\hat{\mathbf{k}}^{-}_{1,2, \perp} \times \hat{\mathbf{z}}) \notag \\
  &\quad+ \frac{A_1 A_3 \,\delta(\mathbf{k - k^-_{1,3}})\, e^{i\eta_{1,3}^-}\,\omega^{-}_{1,3}}
          {c^2|\mathbf{k}_1 - \mathbf{k}_3|^2 - (\omega^-_{1,3})^2}
     \,\vec{\boldsymbol{\beta}}_{1,3} \cdot (\hat{\mathbf{k}}^{-}_{1,3, \perp} \times \hat{\mathbf{z}})
   \Biggr] + \mathrm{c.c.} 
\label{eq:e_2_kperp_cross_z} \\
\hat{E}^{(2), \rm nres}_{\hat{k}_{\perp}} 
  &= -\frac{(2\pi)^3}{4 B_0} \Biggl[
     \frac{A_2 A_3 \,\delta(\mathbf{k - k^+_{2,3}})\, e^{i\eta_{2,3}^+}}
          {c^2 |k_{2,z} + k_{3,z}|^2 - (\omega_{2,3}^+)^2}
     \left(2 k_{2,3,z}^{+} k_{2,3,\perp}^+ c^2 \alpha_{2,3}
     - c \omega^+_{2,3} (\vec{\boldsymbol{\beta}}_{2,3} + \vec{\boldsymbol{\beta}}_{3,2})
       \cdot \hat{\mathbf{k}}^+_{2,3\perp}\right)
   \Biggr] + \mathrm{c.c.} 
\label{eq:e_2_kperp}
\end{align}
The total second-order electric field $\bfm{E^{(2)}}$ that can produce $0$ frequency modes by interacting with linear waves is the integral of Eqs (\ref{eq:e_2_kperp_cross_z})-(\ref{eq:e_2_kperp}) over $k$-space (i.e., the inverse Fourier-transform), combined with terms that oscillate at the linear wave frequency in Eq. (\ref{eq:e2_z}). The second-order magnetic field $\bfm{B^{(2)}}$ can be obtained from the electric field using Faraday's Law. These fields are given by:

\begin{align}
\bfm{E}^{(2)} &= -\frac{1}{4 B_0} \Biggl[ \frac{A_1 A_2 \, e^{i \xi^-_{1,2}} |k_{3,z}|  \,\bfm{\hat{z}} \cdot (\,\vec{\boldsymbol{\beta}}_{1,2} \times \hat{\mathbf{k}}_{3, \perp})}{k_{3,\perp}^2} (\hat{\mathbf{k}}_{3, \perp} \times \hat{\mathbf{z}}) +  \frac{A_1 A_3 \, e^{i \xi^-_{1,3}} |k_{2,z}|  \,\bfm{\hat{z}} \cdot (\,\vec{\boldsymbol{\beta}}_{1,3} \times \hat{\mathbf{k}}_{2, \perp})}{k_{2,\perp}^2} [\hat{\mathbf{k}}_{2, \perp} \times \hat{\mathbf{z}}] \notag \\
&\quad - \frac{A_2 A_3 e^{i \xi^+_{2,3}} (2 k_{1,z} k_{1,\perp} \alpha_{2,3} - k_1 (\vec{\boldsymbol{\beta}}_{2,3} + \vec{\boldsymbol{\beta}}_{3,2})
\cdot \bfm{\hat{k}_{1,\perp}}) }{k_{1,\perp}^2} \bfm{\hat{k}_{1,\perp}} \notag \\
&\quad + \left( A_1 A_2 \alpha_{1,2} e^{i\xi^-_{1,2}} + A_1 A_3 \alpha_{1,3} e^{i\xi^-_{1,3}}  + 2A_2A_3 \alpha_{2,3} e^{i\xi^+_{2,3}}\right) \bfm{\hat{z}}\Biggr] + \mathrm{c.c.} \label{e2_total} \\
\bfm{B}^{(2)} &= -\frac{1}{4 B_0} \Biggl[\frac{A_1 A_2 \, e^{i \xi^-_{1,2}}   \,\bfm{\hat{z}} \cdot (\,\vec{\boldsymbol{\beta}}_{1,2} \times \hat{\mathbf{k}}_{3, \perp})}{k_{3,\perp}^2} (\bfm{k}_{3}  \times (\hat{\mathbf{k}}_{3, \perp} \times \bfm{\hat{z}})) +  \frac{A_1 A_3 \, e^{i \xi^-_{1,3}} \,\bfm{\hat{z}} \cdot  (\,\vec{\boldsymbol{\beta}}_{1,3} \times \hat{\mathbf{k}}_{2, \perp})}{k_{2,\perp}^2} ( \bfm{k_2} \times (\hat{\mathbf{k}}_{2, \perp} \times \bfm{\hat{z}})) \notag \\
&\quad - \frac{A_2 A_3 e^{i \xi^+_{2,3}}(2 k_{1,z} k_{1,\perp} \alpha_{2,3} - k_1 (\vec{\boldsymbol{\beta}}_{2,3} + \vec{\boldsymbol{\beta}}_{3,2})
\cdot \bfm{\hat{k}_{1,\perp}}) }{k_1  k_{1,\perp}^2} (\bfm{k_1} \times \bfm{\hat{k}_{1,\perp}})   \notag \\
&\quad  +  \left( \frac{A_1 A_2 \alpha_{1,2} e^{i\xi^-_{1,2}}}{|k_{3,z}|} (\bfm{k_3} \times \bfm{\hat{z}}) + \frac{A_1 A_3 \alpha_{1,3} e^{i\xi^-_{1,3}}}{|k_{2,z}|} (\bfm{k_2} \times \bfm{\hat{z}})  + \frac{2A_2A_3 \alpha_{2,3} e^{i\xi^+_{2,3}}}{k_1} (\bfm{k_1} \times \bfm{\hat{z}})\right) 
\Biggr] + \mathrm{c.c.} \label{b2_total}
\end{align}

To check whether these fields produce a zero-frequency current only in the $\hat{y}$ direction, we can compute $\bfm{\bar{J}^{(3)}}$ non-linear current using the ansatz $\bfm{E} = \bfm{E^{(1)}} + \bfm{E^{(2)}}$ and $\bfm{B} = B_0 \bfm{\hat{z}} + \bfm{B^{(1)}} + \bfm{B}^{(2)}$ and take its volume average: 

\begin{equation}
    \bfm{\bar{J}^{(3)}} = \frac{1}{V} \int \Biggl[\frac{(\dvg{E} \, (\bfm{E} \times \bfm{B}) - \bfm{B}(\bfm{B} \cdot (\curl{B}) - \bfm{E} \cdot (\curl{E}))}{|\bfm{B}|^2} \Biggr] dV
    \label{eq:j_third_order_avg}
\end{equation}

\begin{figure}
    \centering
    \includegraphics[width=0.9\textwidth]{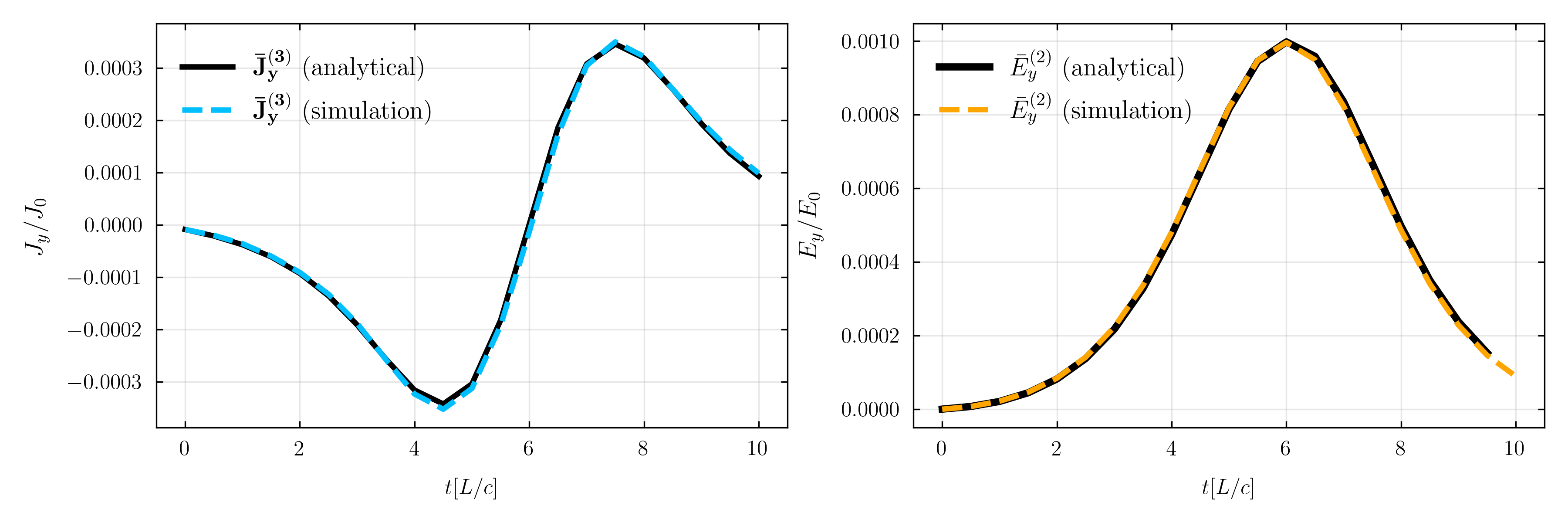}
    \caption{(Left): Comparison of the volume averaged force-free current from the $F \longleftrightarrow A+A$ simulation and analytical results from Eq. (\ref{eq:j_third_order_avg}) as a function of time. The analytical results depend on the values of $A_1, A_2, A_3$ and $\phi_1,\phi_2, \phi_3$, which are extracted from the FFE simulation as a function of time. The simulation current is computed directly using on the full simulation electric and magnetic fields. (Right): Similar comparison between the analytical and simulation electric fields. The analytical electric fields is computed by integrating analytical $\bfm{\bar{J}^{(3)}}$ over time. In both cases, there is an excellent agreement between the simulation values and analytical calculations. 
    }
    \label{fig:j3_e2_comparison}
\end{figure}

The expanded expression for $\bfm{\bar{J}^{(3)}}$ is long and cumbersome to reproduce on the paper. We compute it analytically by plugging in Eqs. (\ref{eq:e_f1})-(\ref{eq:b_a3}) and (\ref{e2_total})-(\ref{b2_total}) into Eq. (\ref{eq:j_third_order_avg}) using the symbolic algebra library \texttt{SYMPY} \citep{sympy_10.7717/peerj-cs.103}. In order to compute the numerical value of  $\bfm{\bar{J}^{(3)}}$, the values of $A_i$ and $\phi_i$ for $i=1,2,3$ are computed directly from the simulation data using Fourier decomposition. Figure \ref{fig:j3_e2_comparison} shows an excellent agreement between the values of $\bfm{\bar{J}}^{(3)}$ and $\bfm{\bar{E}}^{(2)}$ using the analytical calculation and simulation data.

\section{Multi Wave System} \label{appendix:many_wave_equation}
We derive the wave kinetic equations for the case when the waves are randomly phased. Our derivation follows  \cite{Zakharov1992} but for a system that supports multiple kinds of linear waves. It is complementary to the derivation by \citep{golbraikh_2023ApJ...957..102G}. From Eqs. (\ref{eq:Adot})-(\ref{eq:adot}), we can write down the time evolution equations for $|\tilde a_{\bfm{k}_i}|^2$ and $|\tilde A_{\bfm{k}_i}|^2$:
\begin{align}
    \frac{\dd  (\tilde{A}_{\bfm{k_1}} \tilde{A}_{\bfm{k_1}}^*)}{\dd t} &=  \text{Im} \left[ \iint \left(2\tilde{V}^{F}_{123} \tilde{A}_{\bfm{k_2}}^* \tilde{a}_{\bfm{k_3}}^* \tilde{A}_{\bfm{k_1}} +  2\tilde{V}^{F}_{213} \tilde{A}_{\bfm{k_2}}^* \tilde{a}_{\bfm{k_3}} \tilde{A}_{\bfm{k_1}} + \tilde{V}^{A}_{123} \tilde{a}_{\bfm{k_2}}^* \tilde{a}_{\bfm{k_3}}^* \tilde{A}_{\bfm{k_1}} \right) \dd^3\bfm{k_2} \dd^3 \bfm{k_3} \right] \label{eq:Ndot_before} \\
     \frac{\dd  (\tilde{a}_{\bfm{k_1}} \tilde{a}_{\bfm{k_1}}^*)}{\dd t} &=  \text{Im} \left[ \iint \left(2\tilde{V}^{F}_{231} \tilde{A}_{\bfm{k_2}}^* \tilde{A}_{\bfm{k_3}} \tilde{a}_{\bfm{k_1}} +  
     2\tilde{V}^{F}_{231} \tilde{A}_{\bfm{k_2}}^* \tilde{a}_{\bfm{k_3}} \tilde{a}_{\bfm{k_1}}  \right) \dd^3\bfm{k_2} \dd^3 \bfm{k_3} \right] \label{eq:ndot_before}.
\end{align}
Note that in our definitions, $\tilde{V}^{A/F}$ are purely real, and $\tilde{A}_{\bfm{k_i}}$ and $\tilde{a}_{\bfm{k_i}}$ are complex numbers. If the spectrum of waves is randomly phased, then to the lowest order, the time average of all the cubic terms inside the integral is zero. That is, $<\tilde{A}_{\bfm{k2}}^* \tilde{a}^*_{\bfm{k3}} \tilde{A}_{\bfm{k_1}}>, <\tilde{A}_{\bfm{k2}}^* \tilde{a}_{\bfm{k3}} \tilde{A}_{\bfm{k_1}}>, \cdots = 0$, where `$<\cdots>$' denotes the time average over multiple wave periods. This implies that the wave amplitudes $|a_{\bfm{k}_i}|$ evolve on a timescale \textit{slower} than the one estimated in Eq (\ref{eq:first_order_timescale}). To get the relevant timescales, the integral on the RHS of Eqs (\ref{eq:Ndot_before})-(\ref{eq:ndot_before}) needs to be evaluated at a higher order. 
Defining $J^{F}_{231} \delta(\bfm{k}_2 - \bfm{k}_3 - \bfm{k}_1)  = \left\langle\tilde{A}_{\bfm{k_2}} \tilde{a}_{\bfm{k_3}} \tilde{A}_{\bfm{k_1}}^*\right\rangle$, $J^{A}_{231} \delta(\bfm{k}_2 - \bfm{k}_3 - \bfm{k}_1) = \left\langle\tilde{A}_{\bfm{k_2}} \tilde{a}_{\bfm{k_3}}^* \tilde{a}_{\bfm{k_1}}^*\right\rangle$, $N_{\bfm{k}} \delta(\bfm{k} - \bfm{k}_1) = \left\langle \tilde A_{\bfm{k}} \tilde A^*_{\bfm{k}_1} \right\rangle$ and $n_{\bfm{k}} \delta(\bfm{k} - \bfm{k}_1) = \left\langle \tilde a_{\bfm{k}} \tilde a^*_{\bfm{k}_1} \right\rangle$, we can write the time-averaged Eqs. (\ref{eq:Ndot_before})-(\ref{eq:ndot_before}) in terms of $J^{(A/F)}_{ijk}$:
\begin{align}
     \left\langle\frac{{\dd}  N_{\bfm{k}_1}}{\dd t} \right\rangle&=  \text{Im} \left[ \iint \left(2\tilde{V}^{F}_{123} J^{F}_{231} +  2\tilde{V}^{F}_{213} (J^{F}_{132})^* + \tilde{V}^{A}_{123} (J^A_{123})^* \right) \dd^3\bfm{k_2} \dd^3 \bfm{k_3} \right] \label{eq:Ndot_avg} \\
     \left\langle\frac{{\dd} n_{\bfm{k}_1}}{\dd t} \right\rangle &=  \text{Im} \left[ \iint \left(2\tilde{V}^{F}_{231} (J^{F}_{321})^* +
     2\tilde{V}^{A}_{231} J^A_{231}  \right) \dd^3\bfm{k_2} \dd^3 \bfm{k_3} \right] \label{eq:ndot_avg}.
\end{align}
The next-order values of $J^F_{ijk}$ and $J^A_{ijk}$ can be obtained by solving for their governing equations. For example, $J^{F}_{231}$ obeys the following differential equation:
\begin{align}
    \frac{\partial J^{F}_{231}}{\partial t} =   \left\langle \frac{\partial \tilde{A}_{\bfm{k_2}}}{\partial t} \tilde{a}_{\bfm{k_3}} \tilde{A}_{\bfm{k_1}}^* \right\rangle +  \left\langle \tilde{A}_{\bfm{k_2}} \frac{\partial \tilde{a}_{\bfm{k_3}}}{\partial t}\tilde{A}_{\bfm{k_1}}^*  \right\rangle + \left\langle  \tilde{A}_{\bfm{k_2}} \tilde{a}_{\bfm{k_3}}  \frac{\partial \tilde{A}_{\bfm{k_1}}^*}{\partial t} \right\rangle .
\end{align}
Making use of Eqs (\ref{eq:Adot}) and (\ref{eq:adot}), and applying the random phase approximation: $\left\langle\tilde{A}_{\bfm{k_1}} \tilde{A}_{\bfm{k_2}}^* \tilde{a}_{\bfm{k_3}} \tilde{a}_{\bfm{k_4}}^*\right\rangle = N_{\bfm{k_1}}n_{\bfm{k_3}} \delta(\bfm{k_1} - \bfm{k_2}) \delta (\bfm{k_3} - \bfm{k_4})$, $\left\langle\tilde{A}_{\bfm{k_1}} \tilde{A}_{\bfm{k_2}}^* \tilde{A}_{\bfm{k_3}} \tilde{A}_{\bfm{k_4}}^*\right\rangle = N_{\bfm{k_1}}N_{\bfm{k_3}} (\delta(\bfm{k_1} - \bfm{k_2}) \delta (\bfm{k_3} - \bfm{k_4}) + \delta(\bfm{k_1} - \bfm{k_4}) \delta (\bfm{k_3} - \bfm{k_2}))$, etc., we can obtain the solution for $J^{F}_{231}(t)$:
\begin{align}
    J^{F}_{231}(t) = \left\langle J^{F}_{231}(0) e^{i(\omega^A_{\bfm{k_2}} + \omega^{F}_{\bfm{k_3}} - \omega^F_{\bfm{k_1}})t} \right\rangle + \lim_{\gamma\to0}  \frac{\tilde{V}^{F}_{123}\left[N_{\bfm{k_1}} ( n_{\bfm{k_3}} + N_{\bfm{k_2}}) - N_{\bfm{k_2}} n_{\bfm{k_3}} \right]}{(\omega^A_{\bfm{k_2}} + \omega^{F}_{\bfm{k_3}} - \omega^F_{\bfm{k_1}}) + i\gamma}.
\end{align}
In the above solution, we have added a small damping term $\gamma$ to avoid the singularity at the exact resonance. In the large time limit, the time average of the first term goes to $0$, and $J$ can be approximated using the second term. The imaginary part of $J^{F}_{231}$ is given by:
\begin{align}
    \text{Im}[J^{F}_{231}] = -\pi \tilde{V}^{F}_{123}\left[N_{\bfm{k_1}} ( n_{\bfm{k_3}} + N_{\bfm{k_2}}) - N_{\bfm{k_2}} n_{\bfm{k_3}} \right] \delta(\omega^A_{\bfm{k_2}} + \omega^{F}_{\bfm{k_3}} - \omega^F_{\bfm{k_1}})
    \label{eq:jf231}
\end{align}
Similarly, we can write down the solution to the other $J^{A/F}_{ijk}$ terms in Eqs. (\ref{eq:Ndot_avg})-(\ref{eq:ndot_avg}). The imaginary parts of these terms are given by:
\begin{align}
     \text{Im}[(J^{F}_{132})^*] &= \pi \tilde{V}^{F}_{213}\left[N_{\bfm{k_2}} ( n_{\bfm{k_3}} + N_{\bfm{k_1}}) - N_{\bfm{k_1}} n_{\bfm{k_3}} \right] \delta(\omega^A_{\bfm{k_1}} + \omega^{F}_{\bfm{k_3}} - \omega^F_{\bfm{k_2}}) \label{eq:jf132} \\
    \text{Im}[(J^{F}_{321})^*] &=  \pi \tilde{V}^{F}_{231}\left[N_{\bfm{k_1}} ( n_{\bfm{k_2}} + N_{\bfm{k_3}}) - N_{\bfm{k_3}} n_{\bfm{k_2}} \right] \delta(\omega^A_{\bfm{k_3}} + \omega^{F}_{\bfm{k_2}} - \omega^F_{\bfm{k_1}}) \label{eq:jf321} \\
     \text{Im}[J^{A}_{231}] &=  -\pi \tilde{V}^{A}_{231}\left[n_{\bfm{k_3}} n_{\bfm{k_1}} - N_{\bfm{k_2}} (n_{\bfm{k_1}} + n_{\bfm{k_3}}) \right] \delta(\omega^A_{\bfm{k_3}} + \omega^{A}_{\bfm{k_1}} - \omega^F_{\bfm{k_2}}) \label{eq:ja231} \\
     \text{Im}[(J^{A}_{123})^*] &=  \pi \tilde{V}^{A}_{123}\left[n_{\bfm{k_2}} n_{\bfm{k_3}} - N_{\bfm{k_1}} (n_{\bfm{k_3}} + n_{\bfm{k_2}}) \right] \delta(\omega^A_{\bfm{k_2}} + \omega^{A}_{\bfm{k_3}} - \omega^F_{\bfm{k_1}}) \label{eq:ja123} 
\end{align}
Substituting Eqs. (\ref{eq:jf231})-(\ref{eq:ja123}), into Eqs. (\ref{eq:Ndot_avg})-(\ref{eq:ndot_avg}), we finally obtain the wave kinetic equations. Defining  $W^{F}_{ijk} = 2 \pi |\tilde{V}^F_{ijk}|^2 \delta(\omega^F_{\bfm{k_j}} + \omega^{A}_{\bfm{k_k}} - \omega^F_{\bfm{k_i}})$ and $W^{A}_{ijk} = 2 \pi |\tilde{V}^A_{ijk}|^2 \delta(\omega^A_{\bfm{k_j}} + \omega^{A}_{\bfm{k_k}} - \omega^F_{\bfm{k_i}})$,

\begin{align}
\left\langle \frac{\dd N_{\bfm{k_1}}}{\dd t} \right\rangle
&= \iint \Big[
    -W^{F}_{123} \Big( N_{\bfm{k_1}} ( n_{\bfm{k_3}} + N_{\bfm{k_2}}) - N_{\bfm{k_2}} n_{\bfm{k_3}} \Big) \nonumber \\[4pt]
&\quad + W^{F}_{213} \Big( N_{\bfm{k_2}} ( n_{\bfm{k_3}} + N_{\bfm{k_1}}) - N_{\bfm{k_1}} n_{\bfm{k_3}} \Big) \nonumber \\[4pt]
&\quad + \tfrac{1}{2} W^{A}_{123} \Big( n_{\bfm{k_2}} n_{\bfm{k_3}} - N_{\bfm{k_1}} ( n_{\bfm{k_3}} + n_{\bfm{k_2}} ) \Big)
\Big] \, \dd^3 \bfm{k_2} \, \dd^3 \bfm{k_3} \\[8pt]
\left\langle \frac{\dd n_{\bfm{k_1}}}{\dd t} \right\rangle
&= \iint \Big[
    W^{F}_{231} \Big( N_{\bfm{k_1}} ( n_{\bfm{k_2}} + N_{\bfm{k_3}} ) - N_{\bfm{k_3}} n_{\bfm{k_2}} \Big) \nonumber \\[4pt]
&\quad - W^{A}_{231} \Big( n_{\bfm{k_3}} n_{\bfm{k_1}} - N_{\bfm{k_2}} ( n_{\bfm{k_1}} + n_{\bfm{k_3}} ) \Big)
\Big] \, \dd^3 \bfm{k_2} \, \dd^3 \bfm{k_3}
\end{align}
The above equations are in agreement with Eqs (4)-(5) of \cite{golbraikh_2023ApJ...957..102G}. 

\section{Non-resonant Three-wave Interactions} \label{appendix:Non-resonant Three-wave Interactions}
The parametric decay and the merger of two lower frequency modes into a higher frequency one in the $F \longleftrightarrow F+A$ and  $F \longleftrightarrow A+A$ interactions require the resonance conditions from Equations (\ref{eq:resonance_k}) and (\ref{eq:resonance_omega}) to be satisfied. In this section, we explore how a frequency mismatch $\Delta \omega \equiv \omega - \omega_1 - \omega_2$ affects the amplitudes of two modes merging into a high-frequency product mode. Physically, the decorrelation timescale $t_{\rm dec} \sim \frac{2\pi}{\Delta \omega}$ must be longer than the growth timescale $t^{(1)}_{\rm int}$ of the waves to resonantly transfer energy between the modes (e.g., see Eq. (56) of \cite{Li_beloborodov_2019ApJ...881...13L}). We verify this by numerically integrating Equations (\ref{eq:three_wave_rescaled_1})-(\ref{eq:three_wave_rescaled_3}) by including a $e^{i\Delta\omega t}$ term in the matrix element for three different triads of waves with different values of $\Delta \omega$. The wavevectors and initial amplitudes for each of the three triads are given in the Table \ref{tab:non_resonant_simulations}.

\begin{figure*}
    \includegraphics[width=0.33\textwidth]{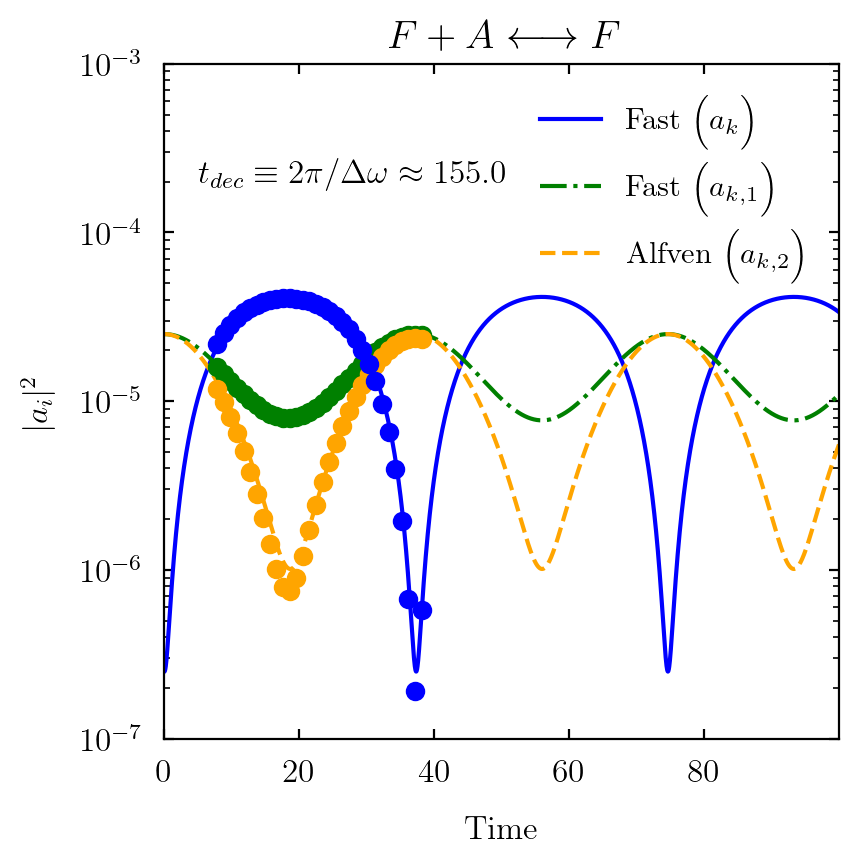}
    \includegraphics[width=0.33\textwidth]{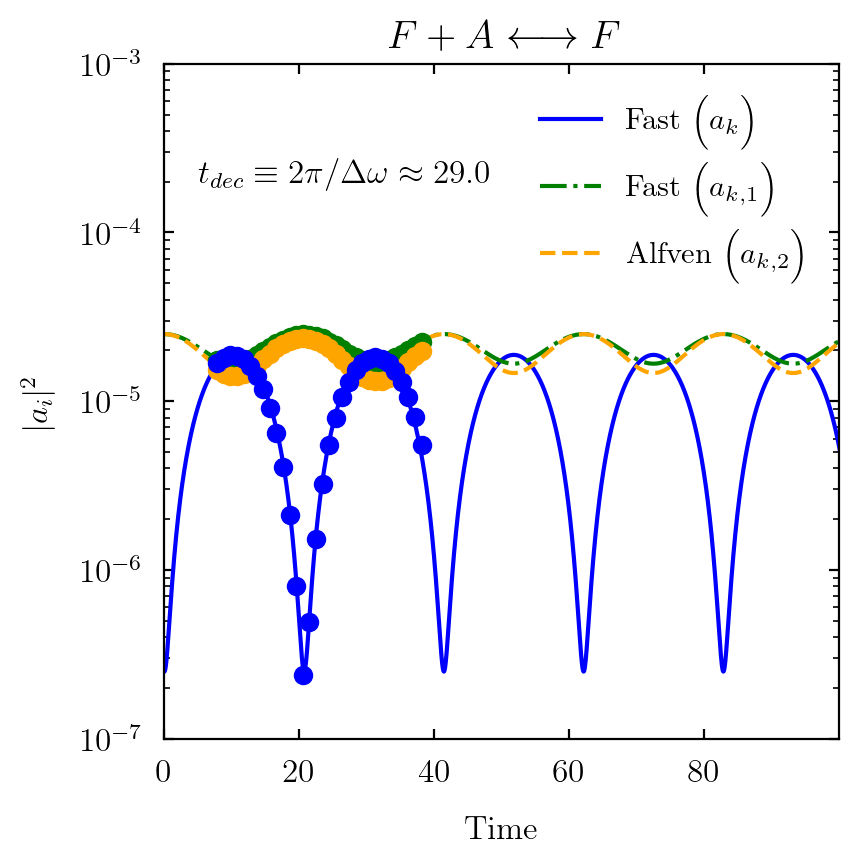}
    \includegraphics[width=0.33\textwidth]{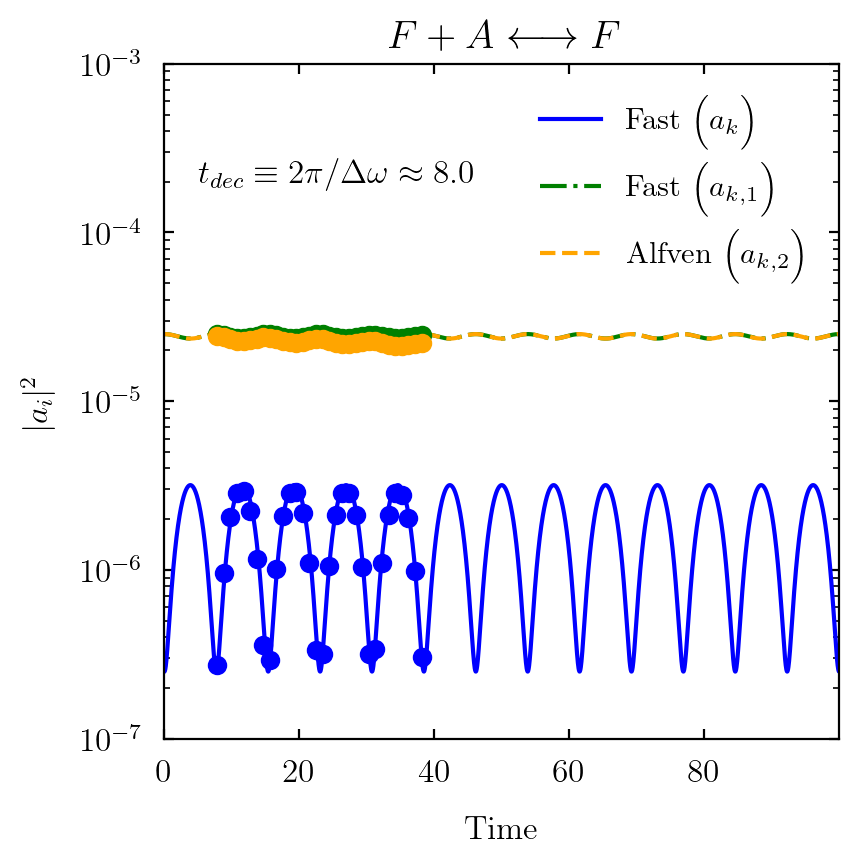}
    \label{fig:non_resonant_three_wave_fig}
    \caption{Numerical solutions to a triad of nearly resonant waves (solid and dashed colored lines) and force-free simulations (colored dots) for the same initial conditions. The initial conditions contain two fast waves and an Alfvén wave, with frequencies $\omega, \omega_1$ and $\omega_2$, respectively, so that they nearly satisfy the resonance conditions for the $FFA$ process. The $y-$axis shows the square of the absolute value of amplitudes, and the $x-$ axis shows the time. The frequency mismatch between the three waves, $\Delta \omega \equiv |\omega - \omega_1 - \omega_2|$, increases from left to right, and as a result, the de-correlation time, $t_{\rm dec}$ decreases. When $t_{\rm dec}$ becomes comparable to the interaction timescale of the three waves, resonant three-wave interactions cease. In the middle and right panels, the high-frequency fast wave no longer grows from the merger of the two lower-frequency waves.}
\end{figure*}

The absolute values of the square wave amplitudes as a function of time are plotted in Figure \ref{fig:non_resonant_three_wave_fig}.  Each color represents $|a_i|^2$, where $a_i$ is the phasor for mode $i$.  The degree of mismatch, i.e., $|\Delta \omega|$, increases from left to right. The solid lines are the numerically integrated solutions, where we use RK-4 integration to advance the time. We also perform force-free simulations for the same set of initial conditions, and the dots represent values taken from the simulation. The simulations are in good agreement with the numerically integrated solutions of the non-resonant three-wave equations.

Since the waves are simulated on a discrete grid, the number of possible wave frequencies is also discretized. This limits the amount of resonant triads of fast and Alfvén waves. Furthermore, the distinct dispersion relations of the fast and Alfvén waves imply that the frequency-matching condition is usually not met. More concretely, the possible wavevectors on the grid are integer multiples of $\frac{2\pi}{L}$, where $L$ is the length of the grid. From the dispersion relation of Alfvén waves, their frequencies on the grid are of the form $\omega = |k_z c| = \frac{2 \pi c |m|}{L}$, where $m \in \mathbb{Z}$ and $c$ is the speed of light. The possible fast wave frequencies instead take the form $\omega = |k c| = \frac{2 \pi c}{L} \sqrt{m_1^2 + m_2^2 + m_3^2}$, with $\{m_1, m_2, m_3\} \in \mathbb{Z}$. The term in the square roots is, in general, not an integer. Therefore, the typical frequency mismatch between a fast wave and a pair of Alfvén waves is $\Delta \omega \sim \frac{2\pi c}{L}$, so that $\frac{\Delta\omega}{\omega} \sim \frac{\lambda}{L}$, where $\lambda$ is the wavelength of the fast wave. Such a triad of waves de-correlates on a timescale of $t_{\rm dec} \equiv \frac{2\pi}{\Delta \omega} \sim \frac{2 \pi L}{\omega \lambda}$. For resonant growth to occur, the decorrelation timescale must be longer than the nonlinear interaction timescale. Using the estimates from \ref{subsection: Interactions of Three Resonant Modes}, this implies that when $\delta B/B_0 \lesssim \frac{\lambda}{2 \pi L}$, modes stop growing resonantly. Tackling the above issue requires a scale separation between the box size, wavelength of interacting waves, and dissipative scale. The former allows resonant interactions to occur, and the latter ensures that there is sufficient phase space for the waves to cascade to smaller scales.

\begin{deluxetable}{ccccccc}
\tablecaption{The columns contain the wavevectors of the product fast and Alfvén waves, the absolute value of frequency mismatch, and the magnetic field amplitudes of the three waves, respectively. The values in the table are used as initial conditions for numerical integration of the three-wave equations and force-free simulations, whose solutions are presented in Figure \ref{fig:non_resonant_three_wave_fig}. The three waves nearly satisfy the $F \longleftrightarrow F+A$ resonance conditions but have an increasing frequency mismatch, going from the top to the bottom row. \label{tab:non_resonant_simulations}}
\tablehead{
\colhead{Triad} & 
\colhead{$k_{a1}$} & 
\colhead{$k_{f1}$} & 
\colhead{$\Delta \omega$} & 
\colhead{$\delta B_{a1} / B_0$} & 
\colhead{$\delta B_{f1} / B_0$} & 
\colhead{$\delta B_{f} / B_0$}
}
\startdata
1 & $\frac{2\pi}{L}(8,8,-10)$ & $\frac{2\pi}{L}(6,0,4)$ & 0.040 & $5 \times 10^{-3}$ & $5 \times 10^{-3}$ & $5 \times 10^{-4}$ \\
2 & $\frac{2\pi}{L}(8,8,-10)$ & $\frac{2\pi}{L}(7,0,4)$ & 0.216 & $5 \times 10^{-3}$ & $5 \times 10^{-3}$ & $5 \times 10^{-4}$ \\
3 & $\frac{2\pi}{L}(8,8,-10)$ & $\frac{2\pi}{L}(9,0,4)$ & 0.785 & $5 \times 10^{-3}$ & $5 \times 10^{-3}$ & $5 \times 10^{-4}$ \\
\enddata
\label{table:resonant_triads}
\end{deluxetable}
\bibliography{main}
\end{document}